\documentclass[usenatbib]{mn2e} \usepackage{psfig}\usepackage{graphicx}\usepackage{amsmath}\usepackage{amssymb}\usepackage{rotating}\usepackage{lscape}

\title[A survey of the Wolf-Rayet stars in NGC\,1313] {A survey of the
  Wolf-Rayet population of the barred, spiral galaxy
  NGC\,1313\thanks{Based on observations made with ESO telescopes at
  the Paranal Observatory under programme ID 076.B-0071 and with
  archival NASA/ESA Hubble Space Telescope, obtained from the
  ESO/ST-ECF Science Archive Facility.}}

\author[L.\,J.\,Hadfield \& P. A. Crowther] {L.\,J.\
  Hadfield\thanks{E-mail: l.hadfield@shef.ac.uk} and P.\,A.\
  Crowther\\ Department of Physics and Astronomy, University of
  Sheffield, Hounsfield Road, Sheffield, S3 7RH, UK\\}

\date{}

\begin{document}
  \maketitle
  
\begin{abstract}   We present a VLT/FORS1 survey of Wolf-Rayet (WR) stars in the
spiral galaxy NGC\,1313.   In total, 94 WR candidate sources have been
identified from narrow-band imaging. Of these, 82 have been
spectroscopically observed, for which WR emission features are
confirmed in 70 cases, one of which also exhibits strong nebular
He\,{\sc ii} $\lambda$4686 emission.  We also detect strong nebular
He\,{\sc ii} $\lambda$4686 emission within two other regions of
NGC\,1313, one of which is a possible supernova remnant. Nebular
properties confirm that NGC\,1313 has a metal-content
log(O/H)+12=8.23$\pm$0.06, in good agreement with previous
studies.  From continuum subtracted H$\alpha$ images we infer a
global star formation rate of 0.6\,M$_{\odot}$yr$^{-1}$. Using
template LMC WR stars, spectroscopy reveals that NGC\,1313 hosts 
a minimum of 84 WR stars.  Our census comprises 51 WN stars,
including a rare WN/C transition star plus 32 WC stars.  In addition,
we identify one WO star which represents the first such case
identified beyond the Local Group. The bright giant H\,{\sc ii} region
PES~1, comparable in H$\alpha$ luminosity to NGC\,595 in M\,33, is
found to host a minimum of 17 WR stars.  The remaining photometric
candidates generally display photometric properties consistent with WN
stars, such that we expect a global WR population of $\sim$115 stars
with N(WR)/N(O)$\sim$0.01 and N(WC)/N(WN)$\sim$0.4.
\end{abstract}


%__________________________________________________________________

\section{Introduction}
\label{introduction}
Wolf-Rayet (WR) stars are evolved, He-burning stars that are descended from
the most massive O stars.  They have a profound influence on their
surroundings since intense stellar winds continually interact with
their local environment returning mechanical energy and processed
material to the ISM.  In recent years, WR stars have received  renewed
interest since they are believed to be the immediate
precursors of Type Ib/c supernovae (SNe) and long $\gamma$-ray
bursts \citep{pac07b}.

WR stars are characterised by strong, broad, emission lines of
nitrogen (WN), carbon (WC) or oxygen (WO).  Their unique spectral
appearance allows WR stars to be readily identified in external galaxies
and have been detected as individual stars in nearby galaxies
\citep[e.g.][]{massey98} and in the integrated starlight of more
distant galaxies \citep{schaerer99}.  WR stars are expected to be
present within instantaneous bursts between 3 and 5 \,Myrs, such that
they trace recent star formation.

%For massive stars, our basic understanding of stellar evolution is
%based on a combination of theoretical and empirical relations which
%are fine-tuned by comparing predictions with observations i.e. the
%number of red to blue to supergiants, WR to O stars and WN to WC
%stars \citep{meynet04}.
%
%Current evolution models fail to reproduce the observed WR populations
%across the full metallicity range.  In response, new stellar evolution
%tracks which address the complexities of stellar rotation have been
%released.  Recently, \citet{vazquez07} demonstrated the fundamental
%role of rotation in massive star evolution and showed that predicted
%stellar populations still failed to reproduce the observed WR statistics
%across a wide metallicity range.  
%In the Local Group WR stars have been well sampled, but most nearby
%dwarf galaxies being metal-poor the observed metallicty range is very
%restricted.  Indeed, M\,31 is the only nearby opportunity to study WR
%populations at solar and super mettalicities. Unfortunately, its large
%extent and unfavourable inclination makes surveying the WR population
%very challenging.

The advent of 8m class telescopes has allowed WR surveys to move
beyond the Local Group, with recent work proving very successful at
locating large numbers of WR stars in nearby ($<$5\,Mpc) galaxies
\citep{schild03, hadfield05}.  Using a combination of narrow-band
imaging sensitive to WR emission features, candidates may be readily
identified in external galaxies, with follow-up spectroscopy
establishing their nature. Consequently, the WR population of several,
nearby galaxies have been studied, increasing the statistics of WR
populations across a broad range of metallicities.  In addition, since WR
stars are expected to be the immediate precursors of SNe and
$\gamma$-ray bursts, it is vital we map the WR
population in nearby galaxies.

%The most
%notable result was that of the metal-rich spiral M\,83 in which more
%than 1000 WR stars were identified \citep{hadfield05}.
%
%Continuing with this approach, our team has examined the WR
%population of the spiral galaxy NGC\,1313 using VLT/FORS1 imaging and
%spectroscopy. 

NGC\,1313 is an isolated, face-on SB(s)d spiral situated at a distance
of 4.1\,Mpc \citep{mendez02}.  With a reported oxygen abundance of
log(O/H) + 12 $\sim$8.3 \citep{walsh97}, the properties of NGC\,1313
are reminiscent of irregular Magellanic-type galaxies, plus 
late-type spirals such as NGC\,300 and M\,33, such that one might expect
NGC\,1313 to host a substantial WR population.  

Previous studies of the stellar content of NGC\,1313 have utilised
high resolution {\it Hubble Space Telescope} ({\it HST}) broad-band photometry
\citep{Larsen04,pellerin07}.  With regards to the WR population, no
direct investigation into the global WR population has been conducted,
although signatures of WR stars have been detected in several bright H\,{\sc
ii} regions within NGC\,1313 \citep{walsh97}.

This paper is structured as follows: the observations and data
reduction techniques used in this analysis are described in
Section~\ref{observations}. In Section~\ref{nebular}, we present
nebular properties for several regions within NGC\,1313.  The WR
population and global content is examined in sections~\ref{WR stars}
and \ref{discussion}. Finally we discuss
and summarise our results in Section~\ref{conclusion} . 

%__________________________________________________________________

\section{Observations and data reduction} 
\label{observations} 

NGC\,1313 has been observed with the ESO Very Large Telescope UT2
(Kueyen) and Focal Reduced/Low dispersion Spectrograph \#1 (FORS1).
The data were obtained using the standard resolution collimator which
covers a 6\arcmin.8 $\times$ 6\arcmin.8\ field of view, with a plate
scale of 0.2\arcsec pixel$^{-1}$. Photometric observations of
NGC\,1313 were obtained on 2004 October 12, with spectroscopic data
following in November 2004 and November--December 2005.  Details of
the observations can be found in Table \ref{ngc1313obs}.

\begin{table}
\caption{VLT/FORS1 observation log for NGC\,1313.}
\begin{tabular}{llcc}
\hline
Date&Filter/&Exposure&DIMM Seeing\\
&Mask ID&(s)&(\arcsec)\\
\hline
\multicolumn{4}{c}{Imaging}\\
2004 October 12&Bessel B&60, 300&0.9\\
&$\lambda$4684, 4871&1500&0.65,0.75\\
&$\lambda$6665, 6563&300& 0.62--0.69\\
\vspace{-0.2cm}\\
\multicolumn{4}{c}{Spectroscopy}\\
2004 November 14&Mask D&3$\times$600&0.6--0.7\\
&Mask E&3$\times$800&0.6--0.8\\
&Mask F&3$\times$800&0.5--0.8\\
2005 November 25&Mask G&3$\times$800&0.7\\
2005 November 27&Mask A&3$\times$300&0.5\\
2005 December 08&Mask B&3$\times$300&0.7--0.8\\
&Mask C&3$\times$600&0.7\\
\hline
\end{tabular}
\label{ngc1313obs}
\end{table}

To supplement our VLT/FORS1 observations, we have retrieved archival
{\it HST} Advanced Camera for Surveys (ACS) Wide Field Camera (WFC)
images of NGC\,1313 obtained with the F435W and F555W filters,  at
three different pointings of NGC\,1313.   All observations comprise a
680s exposure and were obtained for programme ID GO 9774 (PI S.\,S~Larsen).
At the distance of NGC\,1313, the spatial scale of ACS/WFC is
1.0\,pc\,pixel$^{-1}$ (0.05\arcsec\,pixel\,$^{-1}$).

\subsection{Imaging \& Photometry }
\label{vlt:images}

FORS1 was used on the night of 2004 October 12 to obtain narrow-band
images centred on $\lambda 4684$ and $\lambda 4781$ (FWHM=66\AA\ and
68\AA\ respectively).  In addition, narrow-band on- and off-H$\alpha$
images ($\lambda$6563,\,6665\AA, \mbox{FWHM}=61,\,65\AA) were
acquired along with broad-band B images.  All images were taken in
photometric conditions and in seeing conditions of 0.6--0.9\arcsec.

\begin{figure}
\begin{tabular}{c}
\psfig{figure=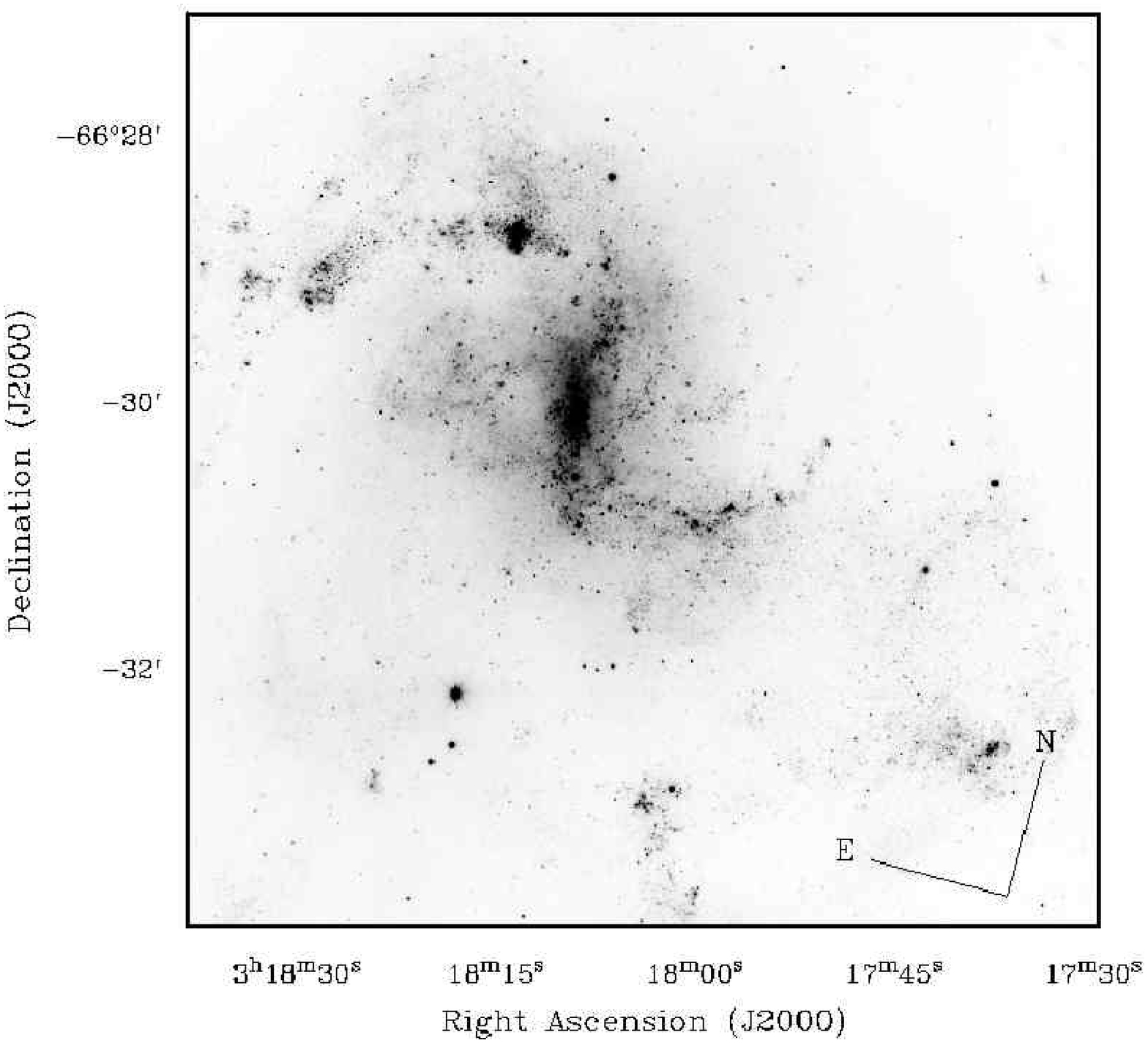,width=\columnwidth,angle=0.}\\
\psfig{figure=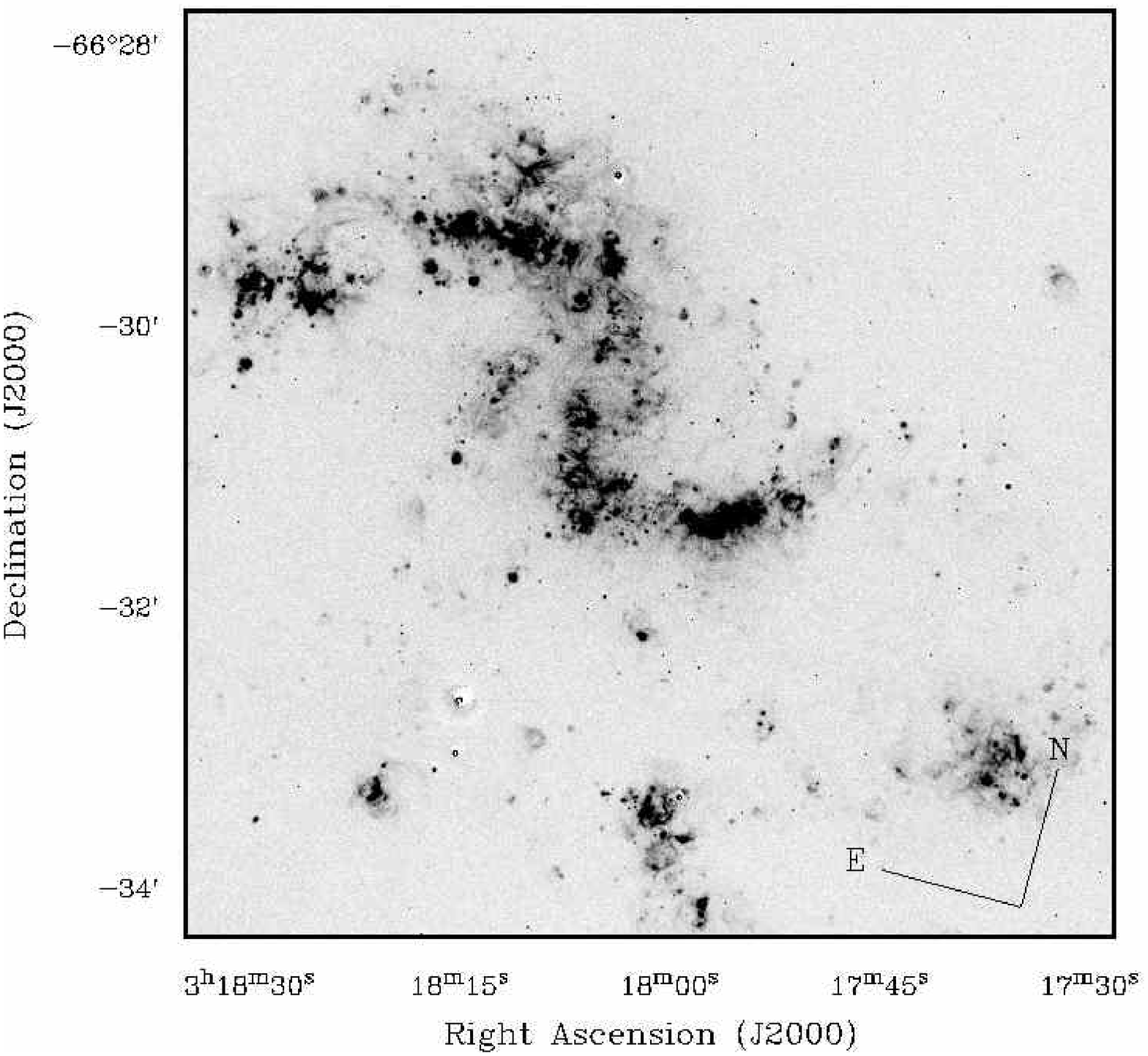,width=\columnwidth,angle=0.}
\end{tabular}
\caption{VLT FORS1 broad-band B (top) and continuum subtracted
  H$\alpha$ (bottom) frames of the surveyed region of NGC\,1313.  The
  orientation of the images is marked and the field of view of each
  frame is 6\arcmin.8 $\times$ 6\arcmin.8. }
\label{fig:images}
\end{figure}

Images were reduced following standard reduction procedures (i.e
debiased, flat field corrected and cosmic ray cleaned) using {\sc
iraf} and {\sc starlink} packages.   

We present broad-band B and continuum subtracted H$\alpha$ images of
the surveyed region in Fig.~\ref{fig:images}, clearly illustrating the
barred-spiral structure of NGC\,1313.  The net H$\alpha$ image of
NGC\,1313 shows that the nebular emission within NGC\,1313 is mainly
concentrated along the spiral arms, with the central bar relatively
faint in ionized gas.  In addition, NGC\,1313 contains three
`satellite' H\,{\sc ii} regions to the south of the nucleus.  In
total, \citet{marcelin83} have catalogued 375 H\,{\sc ii} regions
within NGC\,1313.

Photometry of individual sources within NGC\,1313 was performed using
the {\sc iraf} point-spread (PSF) fitting routine {\sc daophot}.  
\citet{landolt92} photometric standard fields Mark\,A and PG\,2331+055
(containing a total of 7 photometric standard stars) were observed for
absolute calibration of B-band images.  Photometric zeropoints were
found to differ by 0.7\,mag, so we have re-calibrated our photometry
using {\it HST}/ACS F435W images. For the narrow-band images,
photometric zeropoints were obtained by observing the
spectrophotometric standard GD\,71 (m$_{B}$ = 12.8\,mag).

A number of sources do not appear point-like on our VLT/FORS1 images
and have therefore been excluded from our {\sc daophot} photometry.

\subsection{Candidate Selection}

WR candidate stars were identified by searching for He\,{\sc
ii}\,$\lambda$4686/C\,{\sc iii}\,$\lambda$4650 emission (at $\lambda
4684$) relative to the continuum ($\lambda$4781).  As noted by
\citet{hadfield05}, candidate He\,{\sc ii} emission regions are most
readily identified by ``blinking'' the individual $\lambda 4684$ and
$\lambda$4781 images along with a difference image (obtained by
subtracting the $\lambda$4781 frame from the $\lambda 4684$
image).  In total, 94 $\lambda$4684 emission line candidates were
identified in NGC\,1313.  These are listed in the appendix in
Table~A.2.

For the majority of our targets ($\sim$90\%) we have obtained
photometry in at least the $\lambda$4684 filter.  For $\sim$30\% of
these regions it was not possible to obtain a $\lambda$4871 magnitude
since the object was below the detection limit of our photometry. For
five regions PSF photometry was not possible since they were located
in extremely crowded regions of the galaxy. Sources which were not
detected in our images are given an upper limit in Table~A.2.

\subsection{MOS Spectroscopy}

FORS1 Multi Object Spectroscopy (MOS) of candidate WR regions within
NGC\,1313 was undertaken during November 2004 and November--December
2005 in seeing conditions between 0.\arcsec5 and 0\arcsec.8.
Spectra were acquired using a 0\arcsec.8 slit and the 300V grism
(centred on 5850\AA), resulting in a dispersion of 3\AA\,pixel$^{-1}$
and resolution of $\sim$8\AA\ (as measured from comparison arc lines).

In total, 82 of the 94 candidates have been spectroscopically observed
in 7 different MOS mask configurations, which were labelled A to G
(see Table~\ref{observations}).
To maximise S/N, candidates were ranked based on the observed
$\lambda$4684 excess and then grouped according to continuum
brightness. Exposure times ranged between 900s for our brighter
candidates (m$_B$ $\sim$18\,mag) to 2400s for the fainter targets
(m$_B$$\gtrsim$20\,mag). The blue ($\sim$4500\AA) continuum {\it S/N} ranged from
$\sim$50 for the brighter sources to $\lesssim$1 for the very faint
cases. Nevertheless, blue WR features were generally detected at the
20$\sigma$ level.  

Positional restrictions typically permitted $\sim$15 candidates to be
observed simultaneously using FORS1. Since MOS spectroscopy allows the
spectra of 19 targets to be obtained per MOS mask, we have
supplemented our analysis by observing 9 H\,{\sc ii} regions within
NGC\,1313.

For brevity we have assigned \#1--94 and H1--9 nomenclature to our WR
candidates and H\,{\sc ii} regions, respectively. Positional and
photometric information is presented in Tables~A.1
and~A.2 in the appendix.  Additional information
(i.e. spectral classification etc) is included for those regions that
were included in spectroscopic follow-up investigations.  FORS1
narrow-band $\lambda$4684 finding charts can be found in Appendix B.

Data were prepared and processed using standard techniques and {\sc
iraf} and {\sc starlink} packages i.e. bias subtracted, flat field
corrected, extracted and flux / wavelength calibrated, with care being
taken during the extraction process to minimise contamination from
neighbouring sources.  
The wavelength coverage of individual candidates is dependent on their
location on the CCD, but for the majority of targets a spectral
coverage of 3500--7500\AA\ was achieved.

The spectrophotometric standards LTT\,3218 and GD\,108 were observed
in order to relatively flux calibrate the spectra.  Slit-loss
corrections are estimated to be of the order of $\sim$5--20\% for
seeing conditions between 0.5--0.8\arcsec.  FORS1 has an
atmospheric dispersion compensator and so slit-losses will be
independent of wavelength.  Consequently, observed data have been
uniformly scaled by up to 20\% to match the conditions in which they
were obtained.

For very faint cases, no continuum was present on the raw 2D frame
and identification was made solely on the presence of emission
features.  A nearby continuum source was used as a trace during
extraction.

\section{Nebular analysis}
\label{nebular}

Here we derive the nebular properties of several regions within
NGC\,1313.  Of the 82 candidate and 9 H\,{\sc ii} regions
spectroscopically observed, the strength of the [O\,{\sc iii}]
$\lambda$4363 line was detected at the 5$\sigma$ level in only six
regions permitting a determination of T$_{\mbox{e}}$.  These are
NGC\,1313~\#14 and 85, H1, H2, H3 and H5.

The nebular analysis was performed using the {\sc starlink} package
{\sc dipso} with line fluxes being determined using the ELF (emission
line fitting) routine to fit Gaussian line profiles to the spectral
features. Observed and dereddened line fluxes are presented in
Table~\ref{fluxes}. 

\begin{table*}
\caption{Observed (F$_{\lambda}$) and dereddened (I$_{\lambda}$) nebula
  line fluxes of six H\,{\sc ii} regions within NGC\,1313.  Line ratios are
  normalised to H$\beta$ = 100 and we present the observed and
  dereddened H$\beta$ line flux in the penultimate row ($\times
  10^{-13}\mbox{erg}\,\mbox{s}^{-1}\mbox{cm}^{-2}$).}
\begin{tabular}{llrrrrrrrrrrrrr}
\hline
$\lambda_{rest}$&ID&\multicolumn{2}{c}{\#14}&\multicolumn{2}{c}{\#85}&\multicolumn{2}{c}{H1}&\multicolumn{2}{c}{H2}&\multicolumn{2}{c}{H3}&\multicolumn{2}{c}{H4}\\
&&\multicolumn{1}{c}{F$_{\lambda}$}&\multicolumn{1}{c}{I$_{\lambda}$}&\multicolumn{1}{c}{F$_{\lambda}$}&\multicolumn{1}{c}{I$_{\lambda}$}&\multicolumn{1}{c}{F$_{\lambda}$}&\multicolumn{1}{c}{I$_{\lambda}$}&\multicolumn{1}{c}{F$_{\lambda}$}&\multicolumn{1}{c}{I$_{\lambda}$}&\multicolumn{1}{c}{F$_{\lambda}$}&\multicolumn{1}{c}{I$_{\lambda}$}&\multicolumn{1}{c}{F$_{\lambda}$}&\multicolumn{1}{c}{I$_{\lambda}$}\\
\hline
%                    orig.WR85 =>wr14	 orig.WR10=>wr85    orig. H7=>H1      orig. H11=>H2	orig. C4 =>H3   orig. C1=>H4  
3727    &  [O\sc{ii}] & 133.3 & 171.6 &  116.2 &  143.9 &   230.0 &  307.0 &  216.7 &  254.7 &  141.2 &  251.7 & 306.4& 114.4 \\ 
4363    & [O\sc{iii}] &   2.6 &	3.0   &    3.1 &    3.4 &     2.9 &    3.3 &   4.8  &	5.2  &    5.0 &    2.9 &   3.2&   4.5 \\
4861    &    H$\beta$ &   100 &	100   &    100 &    100 &     100 &    100 &   100  &	100  &    100 &    100 &   100&   100 \\
4959    & [O\sc{iii}] & 150.0 & 146.4 &  169.0 &  165.9 &   107.0 &  104.1 &  125.0 &  123.1 &  182.9 &  127.6 & 125.2& 186.6 \\ 
5007    & [O\sc{iii}] & 435.0 & 420.0 &  503.5 &  488.7 &   330.0 &  317.0 &  375.0 &  366.6 &  561.0 &  372.4 & 362.4& 577.5 \\ 
6563    &   H$\alpha$ & 391.7 & 286.0 &  373.4 &  286.0 &   410.0 &  286.0 &  350.0 &  286.0 &  286.0 &  365.5 & 286.0& 371.1 \\ 
6583    &  [N\sc{ii}] &  16.1 &	11.7  &   12.6 &    9.6 &    33.0 &   22.9 &  16.7  &	13.6 &   10.5 &   32.8 &  25.6&  13.7 \\
6717    &  [S\sc{ii}] &    15 &	10.7  &   14.8 &   11.1 &    33.0 &   22.5 &  17.4  &	14.0 &   11.0 &   47.9 &  36.9&  14.5 \\
6731    &  [S\sc{ii}] &  11.2 &	 8.0  &   11.0 &    8.3 &    25.1 &   17.1 &  13.2  &	10.6 &    8.4 &   33.1 &  25.5&  11.1 \\
\smallskip
7330    &  [O\sc{ii}] &   4.4 &	 2.9  &    4.8 &    3.4 &    11.0 &    6.9 &  10.0  &	7.7  &    2.8 &    6.9 &   5.0&   3.9 \\
\smallskip
4861    &    H$\beta$ &  18.0 &	48.1  &   17.3 &   39.8 &     1.0 &    3.0 &   1.2  &	2.3  &   21.9 &    2.9 &   6.2&   9.7 \\
E$_{B-V}$&             & 0.29  &       &   0.25 &        &    0.34 &        &   0.19 &        &   0.23 &        &  0.24&        \\
\hline
\end{tabular}
\label{fluxes}
\end{table*}

\subsection{Interstellar extinction}
\label{exctinction}

The interstellar extinction towards NGC\,1313 has been estimated using
the observed H$\alpha$/H$\beta$ line ratios (accounting for nearby
[N\,{\sc ii}] emission) for the 9 H\,{\sc ii} regions
observed in this analysis (see Table~A.2).
Assuming Case B recombination theory for electron densities of
10$^{2}\,\mbox{cm}^{-3}$ and a temperature of $10^{4}$\,K
\citep{hummer87}, we calculate total $E_{B-V}$ values ranging between
0.19 and 0.56\,mag, with an average $E_{B-V}^{\mbox{{\sc
tot}}}$=0.29$\pm$0.04\,mag.

Underlying H$\alpha$,H$\beta$ absorption from early-type stars is
expected to be of the order of W$_{\lambda}\sim$2\AA. Observed
H$\alpha$ and H$\beta$ equivalent widths are approximately 1550\AA\
and 150\AA, respectively, suggesting that underlying stellar absorption is
negligible.  %However, propagating this component through our
%calculations leads to an additional uncertainty on $E_{B-V}$ of
%$\pm$0.01\,mag.

%As a consistency check we have also estimated $E_{B-V}$ for several
%other regions spectroscopically observed within NGC\,1313.  For the
%majority of the regions,   $E_{B-V}$ was consistent with the mean value
%0.26\,mag.     

Foreground Galactic redding towards NGC\,1313 is estimated to be
E$_{B-V}$=0.10\,mag \citep{schlegel98}, suggesting a modest average
internal extinction of E$_{B-V}$=0.19$\pm$0.05\,mag for NGC\,1313.
All reddening corrections were made following a standard
\citet{seaton79} extinction law and A$_V$/E$_{B-V}$=3.1.

A subset of the regions presented here are associated with the
bright H\,{\sc ii} regions studied by \citet{pagel80} and
\citet{walsh97}. 
We obtain E$_{B-V}$=0.23\,mag for H3, substantially lower than
E$_{B-V}$=c(H$\beta$)/1.64=0.71\,mag derived by \citeauthor{walsh97}
(their No.20). 
Considerable differences are also obtained for the other regions in
common with this study.  Overall, both \citeauthor{pagel80} and
\citeauthor{walsh97} imply a moderately high extinction towards
NGC\,1313, with an average E$_{B-V}$=0.47\,mag
(c.f. E$_{B-V}$=0.29\,mag obtained here). However, a direct comparison
is not straightforward since the specific region sampled in the
different studies is known to differ (i.e. H\,{\sc ii} region No.20 from
\citeauthor{walsh97} is associated with 3 sub-regions observed in our
analysis).

\subsection{Electron densities and temperatures}

Electron densities ({\it N$_{\mbox{e}}$}) and temperatures ({\it
T$_{\mbox{e}}$}) were determined for the six H\,{\sc ii} regions from
diagnostic line ratios presented in Table~\ref{fluxes}, using the
five-level atom calculator {\sc temden}.

Electron densities were calculated assuming a constant {\it
T$_{\mbox{e}}$} of 10\,000K.  For five regions investigated we derive
{\it N$_{\mbox{e}}$}$\sim$100\,cm$^2$, typical of that expected for
H\,{\sc ii} regions.  For H3 a direct determination of
{\it N$_{\mbox{e}}$} was not possible since the [S\,{\sc ii}] 6717/6731\AA\ line ratio
is below the limits of the line diagnostic. Consequently we have
adopted {\it N$_{\mbox{e}}$}=50\,cm$^2$ for temperature and abundance
calculations for this region.

Estimates of the electron temperatures have been derived using the
[O\,{\sc ii}]\,3727/7330\AA\ and [O\,{\sc iii}]\,(4959+5007)/4363\AA\
diagnostic ratios.  For the majority of regions we find that {\it
  T$_{\mbox{e}}$} lies between 9500 and 12\,000K, except for H2 where we
derive a slightly larger electron temperature of $\sim$13\,000K.

%Uncertainties for {\it n$_{e}$} and {\it T$_{e}$} were estimated using
%the formal errors associated with the measured line fluxes. 
Errors on {\it T$_{\mbox{e}}$}(O$^+$) were based on the 10\% measurement
error estimated for the $\lambda$7330\AA\, while we assume that the
15\% formal error associated with [O\,{\sc iii}] 4363\AA\ dominates the
uncertainty on {\it T$_{\mbox{e}}$}(O$^{2+}$).

\subsection{Oxygen Abundance}
\label{abundance}

We have directly measured the O/H content of our six H\,{\sc ii}
regions in NGC\,1313 using the [O\,{\sc ii}] $\lambda$3727 and
[O\,{\sc iii}] $\lambda$5007 nebular emission features along with
associated electron temperatures and densities.   Oxygen
abundances were derived using the {\sc iraf} utility {\sc ionic}. The
total oxygen content for each region was computed assuming that
N(O)/N(H) = N(O$^{+}$)/N(H) + N(O$^{2+}$)/N(H).  Abundance
estimates for individual H\,{\sc ii} regions are presented in
Table~\ref{abundances}.

We derive an average 12 + log(O/H) for NGC\,1313 of 8.23$\pm$0.06.
This agrees with 8.32$\pm0.08$ from \citet{walsh97} to within the
quoted errors. 
Oxygen abundances derived here confirm that NGC\,1313 is somewhat
intermediate in metallicity between the LMC and SMC for which 12 +
log(O/H)$\sim$8.13 and 8.37, respectively \citep{russell90}.

\begin{table}
\caption{Electron temperature, density  \& oxygen abundance for six
  regions within NGC\,1313}
\begin{tabular}{l@{\hspace{1.5mm}}r@{\hspace{1.5mm}}r@{\hspace{1.5mm}}r@{\hspace{1.5mm}}r@{\hspace{1.5mm}}r@{\hspace{1.5mm}}r@{\hspace{1.5mm}}}
\hline
&\multicolumn{6}{c}{Region}\\
Diagnostic&\multicolumn{1}{c}{\#14}&\multicolumn{1}{c}{\#85}&\multicolumn{1}{c}{H1}&\multicolumn{1}{c}{H2}&\multicolumn{1}{c}{H3}&\multicolumn{1}{c}{H4}\\
%Walsh \& Roy\#&\multicolumn{1}{c}{21}&\multicolumn{1}{c}{5}&\multicolumn{1}{c}{23}%&&\multicolumn{1}{c}{20}\\
%Pagel \# & \multicolumn{1}{c}{6} & \multicolumn{1}{c}{3} & & & &  \multicolumn{1}{c}{5} \\ 
\hline
Density&\multicolumn{5}{c}{{\it N}$_{\mbox{e}}$(cm\,$^{-3}$)}\\
 $\lambda$6717/$\lambda$6731            	&80  &80&100	&100	&(50)     &100\\
                                                &$\pm$50 &$\pm$50	&$\pm$60	&$\pm$60	&     &$\pm$60	\\
\\						
\smallskip
Temperature&\multicolumn{5}{c}{{\it T}$_{\mbox{e}}$(K)}\\
$\lambda$3727/$\lambda$7330 			& 9500	  &11\,500&12\,000   &13\,500&11\,000&10\,500\\
                            			&$\pm300$ &$\pm$500&$\pm$500&$\pm$500&$\pm$500&$\pm$400\\
\underline{$\lambda$4959+$\lambda$5007}&10\,500&10\,500&12\,000&13\,000&11\,000&11\,000\\
\multicolumn{1}{c}{$\lambda$4363}   	&$\pm$400 &$\pm$400&$\pm$500&$\pm$600&$\pm$500&$\pm$500\\

\smallskip
Abundances \\
O$^{+}$/H ($\times 10^{-5})$ & 7.2     &  3.0	 & 7.8    &3.1  & 13.4  &  4.5  \\
                                   &$\pm$0.8 &$\pm$0.2   &$\pm$0.9&$\pm$0.1&$\pm$1.6 & $\pm$0.4  \\

O$^{2+}$/H  ($\times 10^{-5})$& 13.5	& 15.8 	& 6.6 	& 5.5 &  8.2  & 14.4  \\
                                  &$\pm$3.9 &  $\pm$4.3	&$\pm$1.5	& $\pm$1.3&  $\pm$ 3.1  & $\pm$4.2 \\
%&$\pm$2.4 &  $\pm$2.1	&$\pm$0.9  	& $\pm$0.6&  $\pm$ 1.3  & $\pm$2.1 \\
O/H ($\times 10^{-5}$)				   & 20.7	& 18.8 	& 14.4	& 8.6  & 21.6  & 18.9  \\
                    &$\pm$4.0 &  $\pm$4.3	&$\pm$1.7	& $\pm$1.3&  $\pm$3.5  & $\pm$4.2 \\
12+log(O/H)			   & 8.32	& 8.27 	& 8.16 	& 7.93  & 8.34  & 8.27  \\
                    &$\pm$0.07&  $\pm$0.08	&$\pm$0.04	& $\pm$0.05&  $\pm$0.06  & $\pm$0.08 \\
Average:&\multicolumn{6}{c}{8.23$\pm0.06$}\\\hline
\end{tabular}
\label{abundances}
\end{table}

Our derived (O/H) abundances show no evidence for a radial dependence
in agreement with previous findings of \citet{walsh97} and
\cite{pagel80}.   This probably results from the barred nature of
the galaxy, in  which the bar provides an effective means of
homogenizing the radial abundance gradients, as demonstrated by
numerical simulations \citep{friedli94}.

\subsection{H$\alpha$ Fluxes}
\begin{table*}
\caption{Dereddened H$\alpha$ fluxes and luminosities for the 12
  primary H\,{\sc ii} regions within NGC\,1313, including the 11
  regions identified by \citet{pagel80} plus one satellite region
  3\arcmin\ to the south-east of the nucleus.  Observed fluxes have
  been derived from continuum subtracted VLT/FORS1 images and have
  been corrected to account for nearby [N\,{\sc ii}]
  emission.  Extinction corrections have been made using the average
  E$_{B-V}$ derived from candidates present within each region.
  Values in parentheses correspond to regions where we adopted our
  average extinction.  In the final row we present the integrated
  H$\alpha$ properties for the galaxy as a whole. Lyman continuum
  ionizing fluxes (Q$_{0}$) have been estimated assuming the
  calibration of \citet{kennicutt98}.}
\begin{tabular}{llllcrrr}
\hline
Region&$\alpha$&$\delta$&Ap&E$_{B-V}$&\multicolumn{1}{c}{{\it F}(H$\alpha$)}&\multicolumn{1}{c}{L(H$\alpha$)}&log Q$_{0}$\\
&\multicolumn{2}{c}{(J2000)}&(\arcsec)&(mag)&\multicolumn{1}{c}{(erg\,s$^{-1}$\,cm$^{-2}$)}&\multicolumn{1}{c}{(erg\,s$^{-1}$)}&(ph\,s$^{-1}$)\\
\hline
        PES\,1&03:18:23.9&-66:28:47.7&20&  0.12&8.7$\times 10^{-13}$&2.3$\times 10^{39}$&51.23\\
        PES\,2&03:18:38.0&-66:29:19.7&12&(0.29)&1.1$\times 10^{-13}$&4.3$\times 10^{38}$&50.50\\
        PES\,3&03:18:37.8&-66:29:34.2&10&  0.25&4.5$\times 10^{-13}$&1.6$\times 10^{39}$&51.07\\
        PES\,4&03:18:16.8&-66:28:44.3&15&  0.26&2.3$\times 10^{-13}$&8.5$\times 10^{38}$&50.79\\
        PES\,5&03:18:05.3&-66:30:27.1& 6&  0.37&3.1$\times 10^{-13}$&1.4$\times 10^{39}$&51.02\\
        PES\,6&03:18:03.2&-66:30:16.5&24&  0.27&6.1$\times 10^{-13}$&2.3$\times 10^{39}$&51.22\\
        PES\,7&03:18:03.5&-66:33:32.3&18&(0.29)&8.4$\times 10^{-14}$&3.3$\times 10^{38}$&50.38\\
        PES\,8&03:17:39.2&-66:31:26.5&18&(0.29)&1.0$\times 10^{-13}$&4.1$\times 10^{38}$&50.47\\
        PES\,9&03:18:18.3&-66:29:02.2&12&(0.29)&6.0$\times 10^{-14}$&2.3$\times 10^{38}$&50.23\\
       PES\,10&03:18:42.2&-66:29:32.0&12&  0.33&1.6$\times 10^{-13}$&6.9$\times 10^{38}$&50.70\\
       PES\,11&03:18:19.6&-66:28:43.5&22&(0.29)&1.7$\times 10^{-13}$&6.5$\times 10^{38}$&50.68\\
\smallskip       
	 P\#12&03:18:23.7&-66:32:55.9&20&0.29&0.9$\times 10^{-13}$&3.7$\times 10^{38}$&50.43\\
NGC\,1313&03:18:14.9&-66:29:33.1&460&(0.29)&2.2$\times 10^{-11}$&8.5$\times 10^{40}$&52.79\\
\hline
\end{tabular}
\label{halpha}
\end{table*}
We have measured net H$\alpha$ fluxes for the 11 bright H\,{\sc ii}
emission regions of NGC\,1313 identified by \citet{pagel80} (denoted
PES\,1--11) plus the `satellite' region 3\arcmin\ to the South-East of the
nucleus (designated P\#12 in Table~\ref{halpha}).  %In addition, we
%have derived the integrated
%H$\alpha$ properties for the entire galaxy.

Narrow-band H$\alpha$ observations of galaxies suffer from
contamination by nearby [N\,{\sc ii}]~$\lambda\lambda$6548,6583
emission. % and global corrections for [N\,{\sc ii}] contamination have
% been well studied and 
For metal-rich spiral galaxies [N\,{\sc ii}] is expected to contribute
1/3 of H$\alpha$ flux, decreasing to only 8\% for metal-poor
irregulars \citep{kennicutt83}.  Indeed, for the six regions presented
in Table~\ref{fluxes}, we estimate [N\,{\sc ii}]/H$\alpha$$\sim$0.06
(assuming [N\,{\sc ii}]\,6548/N\,{\sc ii}]\,6583 $\sim$3).
Consequently, continuum subtracted H$\alpha$ fluxes have been
corrected downwards by 6\%.

Where possible, we have derived an average extinction for these 12
bright H\,{\sc ii} regions using spectroscopy of individual sources in
common with this study. For PES\,7, 8 and 12, this was not possible since
our spectra started redward of H$\beta$.  For PES\,2, 9 and 11 no
spectra were available. For these regions, and for the galaxy as a
whole, we have adopted our average total extinction of
E$_{B-V}$=0.29\,mag.
  
For our assumed distance of 4.1\,Mpc, we derive H$\alpha$ luminosities
for the bright H\,{\sc ii} regions in the range of 2.3$\times 10^{38}$
to $2.3\times 10^{39}$\,erg\,s$^{-1}$. The brightest region (PES\,1)
has an ionizing flux comparable to that observed for NGC\,595 in
M\,33, while the faintest region (PES\,9) is more consistent with that
observed for SMC N19 \citep{kennicutt84}.
For the entire galaxy we derive a H$\alpha$ luminosity of 8.4$\times
10^{40}$\,erg\,s$^{-1}$ which is comparable to $\sim$6500 O7V star
equivalents, assuming a typical O7V star in the SMC has an ionizing
output of 10$^{49}$\,erg\,s$^{-1}$ \citep{hadfield06}.

Of course, the O stars in NGC\,1313 will not be restricted to O7V
spectral types, but will cover the entire spectral range.  For
continuous star formation it is hard to quantify the specific
spectral-type distribution, but it is likely that the actual O star
content of NGC\,1313 is a factor of two higher i.e. 13\,000 stars.  Indeed, the
star formation rate (SFR) in the SMC implies $\sim$400 O7V equivalents
\citep{kennicutt95}, yet the global star content may be as high as
$\sim$1000 \citep{pac07b}.

Using the H$\alpha$--SFR relation of \citet{kennicutt98} we obtain a
global SFR of 0.6\,M$_{\odot}$\,yr$^{-1}$, substantially lower than
the previous H$\alpha$ estimate of 1.4\,M$_{\odot}$\,yr$^{-1}$
\citep{ryder94}.

\subsection{Nebular He\,{\sc ii} $\lambda$4686}
\label{neb_heii}

Except for Planetary Nebulae, nebular He\,{\sc ii} $\lambda4686$
emission is not expected to be observed in ``normal'' H\,{\sc ii}
regions since O stars do not emit sufficient extreme UV photons
($\lambda <$228\AA) to produce a significant He$^+$ ionizing
flux. However, seven such H\,{\sc ii} regions have been identified
within the Local Group \citep{garnett91}, five of which are associated
with WR stars.  Here, we have identified three regions within
NGC\,1313 which exhibit strong nebular He\,{\sc ii} $\lambda4686$
consistent with highly ionized nebulae.

\begin{figure}
\psfig{figure=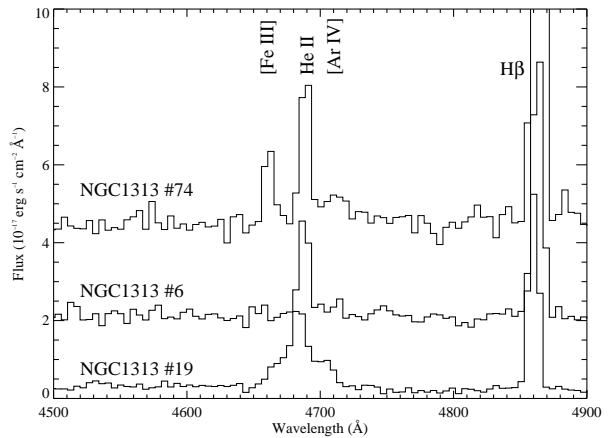,width=\columnwidth,angle=90.}
\caption{Dereddened (E$_{B-V}$=0.29\,mag) VLT/FORS1 spectra of
  NGC\,1313 \#6, 19 and 74, showing the presence of strong, narrow
  nebular He\,{\sc ii} $\lambda$4686.  A broad WR He\,{\sc ii}
  emission is clearly evident in NGC\,1313~\#19.  For clarity each
  spectrum has been successively offset by 1.5$\times
  10^{-17}$erg\,s$^{-1}$\,cm$^{-2}$\AA$^{-1}$.}
\label{neb}
\end{figure}

The dereddened spectra of NGC\,1313~\#6, 19 and 74 are presented in
Fig.~\ref{neb}.  The broad stellar component and superimposed nebular
He\,{\sc ii} $\lambda 4686$ are clearly evident in \#19, as is the presence of
strong [Fe\,{\sc iii}]\,$\lambda$4658 and [Ar\,{\sc
iv}]\,$\lambda$4711 in \#74.  Table~\ref{table:nebheii} lists derived
properties of these three nebulae, along with those for known He\,{\sc
ii} $\lambda$4686 nebulae in the Local Group.
\begin{table*}
\caption{Summary of properties of NGC\,1313~\#6, 19 and 74 and
  He\,{\sc ii} nebulae in the Local Group.  Local He\,{\sc ii} have
  generally been spatially resolved into several components and so we
  list the full range of observed properties.  The large [S\,{\sc
  ii}]/H$\alpha$ observed for N79 corresponds to the supernova remnant
  within the region.\label{table:nebheii}}
\begin{tabular}{l@{\hspace{2.5mm}}l@{\hspace{2.5mm}}l@{\hspace{2.5mm}}l@{\hspace{2.5mm}}c@{\hspace{2.5mm}}c@{\hspace{2.5mm}}cc@{\hspace{2.5mm}}}
\hline
Galaxy&Region&Ionizing&Sp&I($\lambda$4686)/&[S\,{\sc ii}]/& [N\,{\sc ii}] /& Ref\\
&&Source&Type&I(H$\beta$)&H$\alpha$&H$\alpha$\\
\hline
IC1613 &S3 & & WO3 & 0.23 & 0.05 & 0.02 & 2\\
NGC\,1313&\#19&&WN2--4&0.25&0.16&0.08&7\\
SMC&N\,76&AB7&WN4+O6I(f)&0.16-0.24&0.02-0.14&0.02--0.05&5\\
LMC&N79&BAT99-2 &WN2b(h)&0.21--0.56&0.10--0.25 & 0.05--0.26&6\\  
LMC&N206&BAT99-49&WN4:b+O8V&0.08-0.10&0.09-0.22&0.08--0.13&5\\
MW&G2.4\,+\,1.4&WR102&WO2&0.4--1.2&0.08--0.8&0.5--1.0 &3\\
\\
NGC\,1313&\#74&&SNR&0.11&0.50&0.92&7\\
\\
NGC\,1313&\#6&&&0.27&0.17&0.07&7\\
LMC&N159& X-1&HMXB&0.05&&0.24&1\\
LMC&N44C &X-5? &X-ray Neb?&0.02--0.14 & 0.06-0.26 & 0.03--0.35 & 4\\
\hline
\multicolumn{8}{l}{References.---(1)\,\citet{garnett91};
  (2)\,\citet{kingsburgh95b}; (3)\,\citet{Polcaro95}; }\\
\multicolumn{8}{l}{ (4)\,\citet{garnett00}; (5)\,\citet{naze03}; (6)\,\citet{naze03b}; (7)\,this work.}\\
\end{tabular}
\end{table*}

The presence of a WN star within the spectrum of \#19 would suggest
that this is responsible for the hard ionizing photons, and based on
the He\,{\sc ii}\,$\lambda$4686 stellar line flux we infer the presence of
a  single WN2--4 star (see Sect.~\ref{WR stars}).  %We estimate an
%absolute B-band magnitude of -4.2\,mag  for HC\,69, comparable to that
%observed for isolated WN stars.   
We derive a nebular I($\lambda$4686)/I(H$\beta$)=0.25, which is
consistent with that observed for other metal-poor early-type WN
stars \citep[e.g N79 in the LMC;][]{naze03b}.  

For the two regions in which WR stars were not detected, supernova
remnants (SNRs) or X-ray sources are possible ionizing sources.  In
general, the distinction between photoionized and shocked regions is made
using the [S\,{\sc ii}]/H$\alpha$ ratio, with the dividing point being
an empirical value based on observations of known SNRs. Typical
H\,{\sc ii} regions have an observed [S\,{\sc ii}]/H$\alpha$ ratio of
$\sim$0.1 versus 0.5--1.0 observed in SNRs \citep{smith93}.  A
[S\,{\sc ii}]/H$\alpha$=0.4 is generally taken as the
standard discriminator and is found to be independent of metallicity.

\#74 appears to be consistent with known SNR since [S\,{\sc
ii}]/H$\alpha$=0.5.  In addition, [O\,{\sc i}]
$\lambda\lambda$6300,6364 and [N\,{\sc ii}]$\lambda\lambda$6548,6583
are strong with respect to H$\alpha$ also suggesting shock
ionization. We therefore believe that \#74 is a shocked SNR, instead
of a photoionized H\,{\sc ii} region.

For \#6, both [S\,{\sc ii}] and [N\,{\sc ii}] are observed to be much
weaker than in \#74, and with [S\,{\sc ii}]/H$\alpha$=0.2 it is not
consistent with a SNR.  The absence of a hot WR star suggests possible
similarities with N44C and N159 in the LMC, where the high excitation
is linked to possible X-ray sources within the nebulae.  To date,
however, no bright (L$_{X} \geq 10^{38}$erg\,s$^{-1}$) X-ray
counterpart has been linked to \#6 \citep{colbert95, miller98}.

\section{Spectroscopic Results}
\label{WR stars}

Of the 82 candidate regions spectroscopically observed we have
identified broad WR emission features in 70 cases.
Of the 12 cases where we did not detect broad WR emission, 
two sources displayed strong nebular He\,{\sc ii} $\lambda$4686 (see
Sect.~\ref{neb_heii}) and in four cases, the spectral coverage started
longward of the He\,{\sc ii} $\lambda$4686 feature required for WN
classification.  Only six spectra showed no evidence for WR features,
of which three were consistent with foreground late-type stars.

Prior to the present study, the WR population of NGC\,1313 has not
been directly investigated. Strong WR signatures have been detected in
two of the H\,{\sc ii} regions investigated by \citet{walsh97}, their
regions\,3 and 28.  Here, we can confirm the presence of WR stars in both
these regions, although a quantitative comparison is not possible.

Optical spectral classification of WR stars is generally
straightforward, since line diagnostic schemes are well defined
\citep{pac98,smith96}.  Indeed, for most regions distinguishing
between WC (strong C\,{\sc iv} $\lambda$5808), WN (strong He\,{\sc
ii} $\lambda$4686) and WO (strong O\,{\sc vi} $\lambda$3820)
type stars was relatively straightforward. 

For a refined classification of WN stars, we refer to \citet{pac06}
who showed that Magellanic Cloud WN stars could be
further divided into early (WN2--4), mid (WN5--6) and late (WN7--9)
subtypes.
We shall now discuss the WR content of NGC\,1313 in detail.

\subsection{WN population}

In 44 spectra the emission line features present are characteristic of
nitrogen-rich (WN) WR  stars,  with strong He\,{\sc ii}\,$\lambda$4686.
We have assigned a WN2--4 (WNE) or WN7--9 (WNL) subtype to each region
if He\,{\sc ii} $\lambda$4686 is accompanied by N\,{\sc v}
$\lambda$4603--20 or N\,{\sc iii} $\lambda$4634--41, respectively.
For cases where nitrogen emission was not detected, WNE subtypes were
preferred for regions in which broad He\,{\sc ii} $\lambda 4686$
(FWHM$>$20\AA) was observed, otherwise a WNL subtype generally was
assumed.  WN5--6 stars, in which N\,{\sc v} and N\,{\sc iii}
emission is weak, are found to be exclusive to the cores of luminous
H\,{\sc ii} regions within the LMC (e.g. 30 Doradus).  NGC\,1313
contains several such regions, to which we tentatively assign WN5--6
subtypes.  Since these classifications cannot be spectroscopically
confirmed, we note that the WN distribution of NGC\,1313 may be
atypical.  Therefore, to reflect any possible ambiguity surrounding
the classification of these regions, we have adopted WN5--6? subtypes
in Table~A.2.

%
%Based on this criteria, we identify 22 WNE and 22 WNL sources within NGC\,1313.
\begin{figure}
\begin{tabular}{c}
\psfig{figure=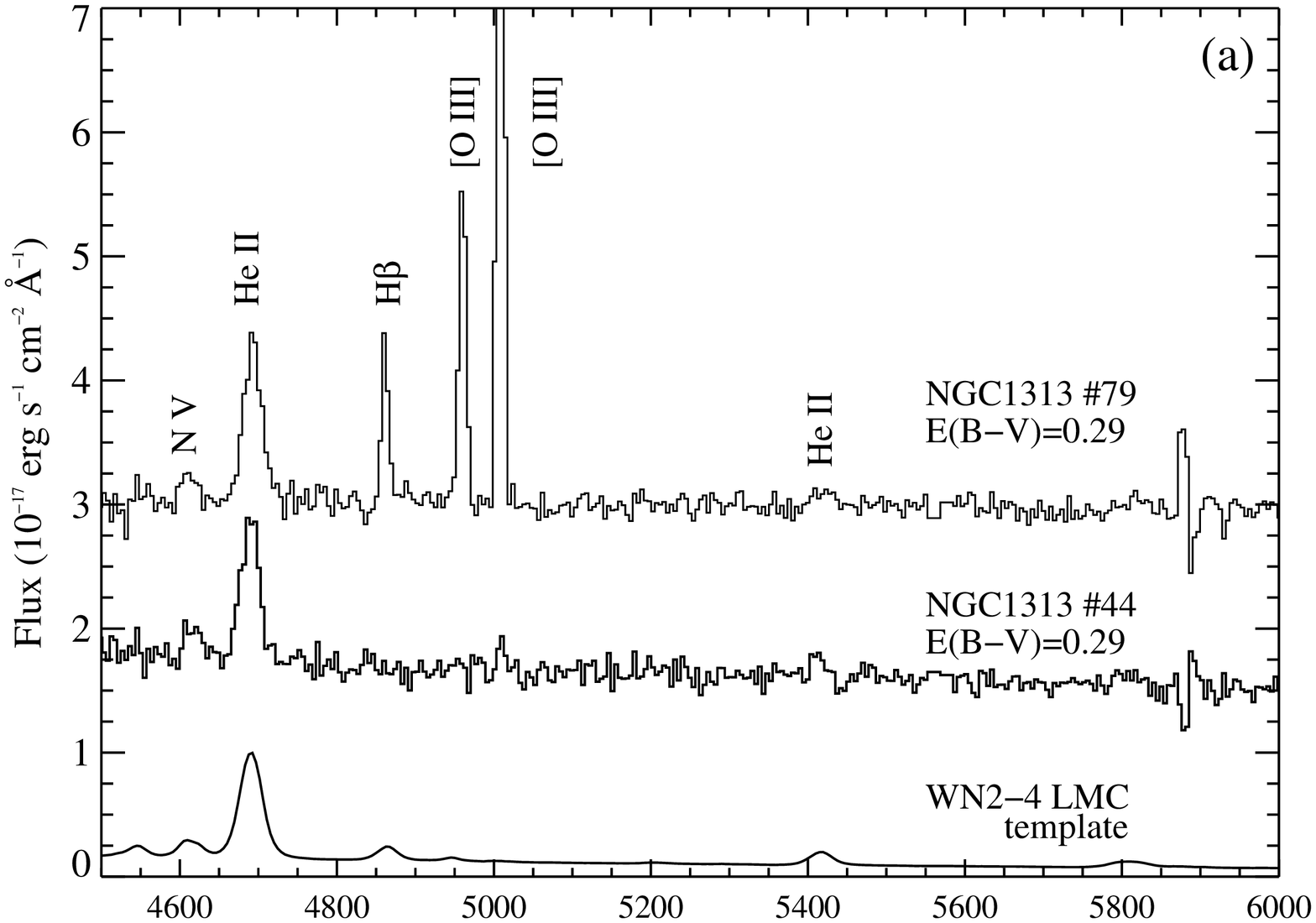,width=\columnwidth,angle=90.}\\
\psfig{figure=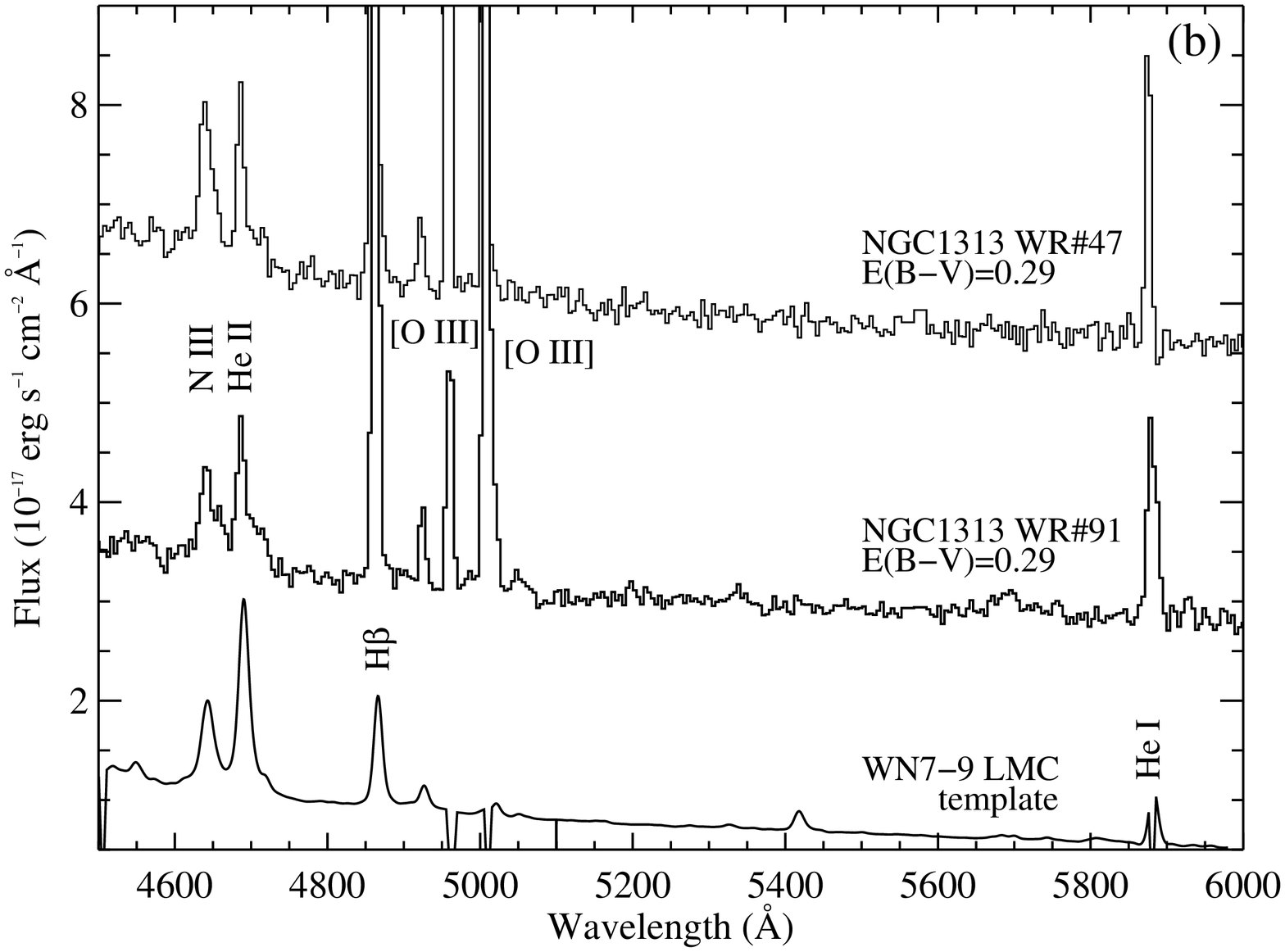,width=\columnwidth,angle=90.}
\end{tabular}
\caption{Dereddened spectral comparison between (a) early and (b) late
  WN regions in NGC\,1313 and ``generic'' optical spectra for single
  WN stars in the LMC \citep{pac06}.  For clarity individual spectra have been
  offset by 2$\times 10^{-17}$erg\,s$^{-1}$\,cm$^{-2}$.}
\label{wn-comp}
\end{figure}

%In both the SMC and LMC, WN2--4 stars dominate the WN
%population.  Since NGC\,1313 shares many similarities with the LMC and
%SMC, it is not unreasonable to expect a similar distribution and that
%the majority of WNE stars within NGC\,1313 are WN2--4 stars.

% 
%The aim of this analysis was to characterise the WR population in
%NGC\,1313. Consequently, we have attempted to estimate the total
%number of WN stars in each object by making comparisons with local
%analogues.   

WN stars are known to exhibit a wide range of intrinsic properties
\citep[e.g.][]{sv98, pac06} and converting between line luminosity and
WR content leads to large uncertainties on derived
populations. Consequently, we suggest a single WN star for each source
unless the line luminosity formally suggests the presence of three or
more stars.

\begin{table}
\caption{Global WR population of NGC\,1313 inferred from spectroscopy
  using dereddened line fluxes and assuming a distance of 4.1\,Mpc.
  The number of WR stars in each source has been estimated using the
  line luminosity calibrations, Log L$_{\rm LMC}$ (erg\,s$^{-1}$), of \citet*{pac06} for
  LMC WR stars. The total inferred WR population adds up to more than the
  number of regions observed since several regions are believed to
  contain multiple WR stars.}
\begin{tabular}{l@{\hspace{2.5mm}}r@{\hspace{2.5mm}}r@{\hspace{2.5mm}}r@{\hspace{2.5mm}}r@{\hspace{2.5mm}}r@{\hspace{2.5mm}}r}
\hline
Subtype&WN2--4&WN5--6?&WN7--9&WN/C&WC4--5&WO\\
Line & $\lambda$4686&$\lambda$4686&$\lambda$4686&$\lambda$4686&$\lambda$5808&$\lambda$5808\\
Log L$_{\rm LMC}$  & 35.92 &36.2&35.86&35.92&36.53&36.01\\
\hline
N(WR)&22&6&22&1&32&1\\
\hline
\end{tabular}
\label{WR:pop}
\end{table}

For the majority of our WN sources, dereddened line fluxes are
consistent with single LMC-like counterparts, assuming average
He\,{\sc ii} $\lambda$4686 line luminosity calibrations of
\citet{pac06} (see Table~\ref{WR:pop}).
%84$\pm71
%\times 10^{35}$erg/s (WN2--4), 175$\pm166 \times 10^{35}$erg/s
%(WN5--6) and 72$\pm67 \times 10^{35}$erg/s (WN7--9) \citep{pac06}.
The exceptions are sources NGC\,1313~\#33 and 58 where we estimate
that 3 WN5--6 and 3 WN2--4 stars, respectively, are required to
reproduce the observed He\,{\sc ii} line flux.

In Fig.~\ref{wn-comp}(a) we compare early WN sources in NGC\,1313 with
a generic LMC WN2--4 template spectrum.  The spectra shown are in
general representative of the observed WNE stars and based on a spectral
comparison we confirm that LMC WN2--4 templates are appropriate
local analogues for WNE stars in NGC\,1313.      
A similar comparison between late-type WN stars in NGC\,1313 and a
generic LMC WN7--9 template star is shown in Fig.~\ref{wn-comp}(b).

From spectroscopy we estimate a total WN population of 50 stars.  The
subtype distribution is summarised in Table~\ref{WR:pop}. For
NGC\,1313~\#1 we quote a WR population of $\geq$1WNE star in
Table~A.2 since it was not possible to extract a reliable spectrum as the object
was located at the edge of the slit. Broad He\,{\sc ii} is clearly
present in the raw 2D image and with no evidence of WC features we
identify a dominant WNE population.

\subsection{WN/C stars} 

For one region, NGC\,1313~\#11, the observed spectral features are not
consistent with normal WN or WC stars.
\begin{figure}
\centerline{\psfig{figure=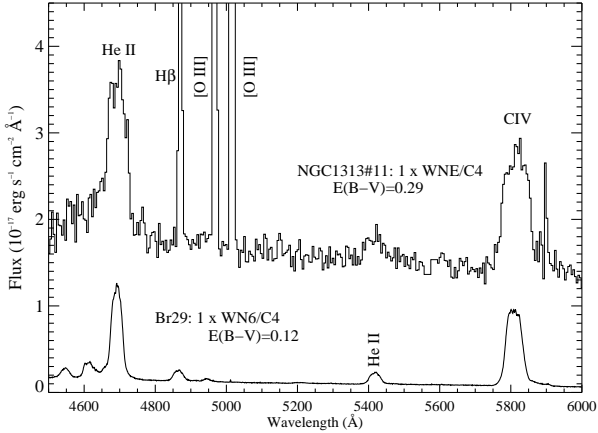,width=\columnwidth,angle=90.}}
\caption{Dereddened spectroscopic comparison between NGC\,1313~\#11
  and LMC WN/C star Brey\,29 (BAT99-36). The spectrum of Brey\,29 is taken
  from \citet{pac95} and has been corrected for an extinction of
  E$_{B-V}$=0.12\,mag and scaled to the distance of NGC\,1313. }
\label{wn/wc}
\end{figure}
In Fig.~\ref{wn/wc} we compare the dereddened FORS1 spectrum of \#11
along with that of the LMC WN5--6/C4 star Brey\,29 (BAT99-36) from
\citet{pac95}.  Strong WC (C\,{\sc iv} $\lambda$5808) and WN (He\,{\sc
ii} $\lambda$4686, $\lambda$5411) features are clearly present in both
stars.
C{\sc iii} $\lambda$4650 is very weak and the relative strength of the
N\,{\sc iii}\,$\lambda$4634-40 and He\,{\sc ii} $\lambda$4686 suggests
a WN5--6 classification, while strong C\,{\sc iv} $\lambda$5808 suggests a
WC4--5 classification.  The similarity in spectral morphology to Brey\,29
suggests a composite WN5--6/C4--5 classification for this source.

Transition WN/C stars are rare, and to date only six such stars have
been identified within the Galaxy, plus two in the LMC.  Since the
total WR populations of the Galaxy and the LMC are 313
\citep{derhucht06, hadfield07} and 134, respectively, we find that the
total observed fraction of WN/C stars for each galaxy is $\sim$1--2\%.
Indeed, this is comparable to observed statistics for NGC\,1313
i.e. N(WN/C)/N(WR)=1/84$\sim$1\%.

\subsection{The WC Content}
\label{wcstars}

\begin{figure}
\centerline{\psfig{figure=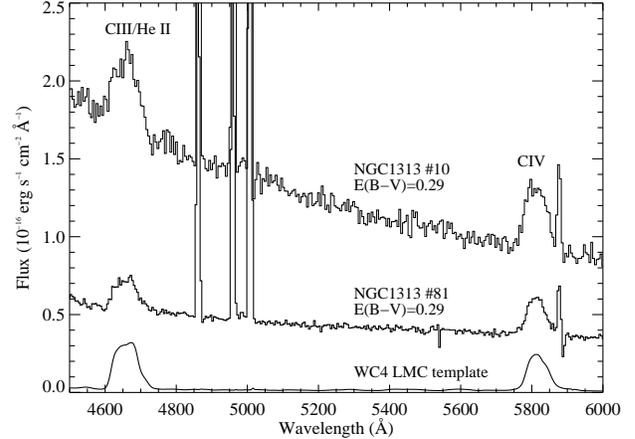,width=\columnwidth,angle=90.}}
\caption{Dereddened spectral comparison between WC regions within
 NGC\,1313 and ``generic'' optical spectra for a LMC-like WC4 star
 \citep{pac06}.  For clarity individual spectra have been offset by
 0.1$\times 10^{-16}$erg\,s$^{-1}$\,cm$^{-2}$\AA$^{-1}$.}
\label{wc-comp}
\end{figure}

For 24 regions spectroscopically observed we have identified broad
blue and red emission features, consistent with those expected from
carbon-rich WR stars.
C\,{\sc iii} $\lambda$5696 is very weak or absent in all 24 spectra
whereas C\,{\sc iv} $\lambda$5808 is present at the 20$\sigma$
level, as illustrated in Fig.~\ref{wc-comp}.  Since WC stars are
classified using the C\,{\sc iii} $\lambda$5696/C\,{\sc iv}
$\lambda$5808 line ratio \citep{smith90}, the absence of C\,{\sc iii}
$\lambda$5696 suggests a WC4--5 subtype for all 24 regions.

For three regions, spectral coverage started longward of
$\lambda$5000 and WC classification was assigned based solely on
the C\,{\sc iv} $\lambda$5808 detection.  For the majority of
sources, dereddened C\,{\sc iv} $\lambda$5808 line luminosities
are consistent with a single LMC-like WC4 star.  For NGC\,1313~\#10,
48, 66 and 88 the observed line fluxes suggest the presence of 2 WC
stars whereas we estimate a content of $\sim$3 WC4--5 stars for
NGC\,1313~\#61 and 64.

\begin{figure}
   \psfig{figure=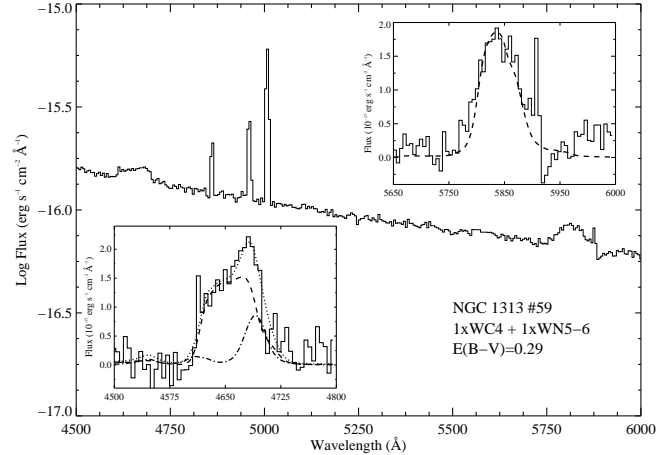,width=\columnwidth,angle=90.}
\caption{Dereddened, velocity-corrected VLT/FORS1 optical
  spectrum of NGC\,1313~\#59.  Also shown are continuum subtracted
  spectral comparisons between the observed and generic blue and red
  WR features.  Generic spectra are taken from \citet{pac06} and have
  been scaled to the distance of NGC\,1313, contributions expected
  from the WC4 (dashed) and WN5--6 (dotted) components are marked.}
\label{fig:mixed}
\end{figure}

For regions NGC\,1313~\#59 and 67, we find that the blue WR features cannot be
reproduced by a population comprising solely of WC stars.
In Fig.~\ref{fig:mixed} we compare the dereddened, continuum subtracted
C\,{\sc iv} $\lambda$5808 profile for \#59 with that expected from a
LMC-like, generic WC4 star at the distance of NGC\,1313.  The observed
red feature is well reproduced by a single WC4--5 star.  In the blue, a
single generic WC4 star cannot reproduce the observed morphology and a
WN5--6 contribution is required.  The
composite spectrum expected from a mixed population reproduces the
observed features exceptionally well, such that we assign a population
of 1\,WC4 and 1\,WN5--6 to \#59.  Similar results are obtained for
NGC\,1313~\#67.
 
\subsection{NGC\,1313~\#31 - The first WO star beyond the Local Group}

In addition to identifying WN and WC stars within NGC\,1313, our WR
survey has revealed the signature of a WO star in one case. 

\begin{figure}
\centerline{\psfig{figure=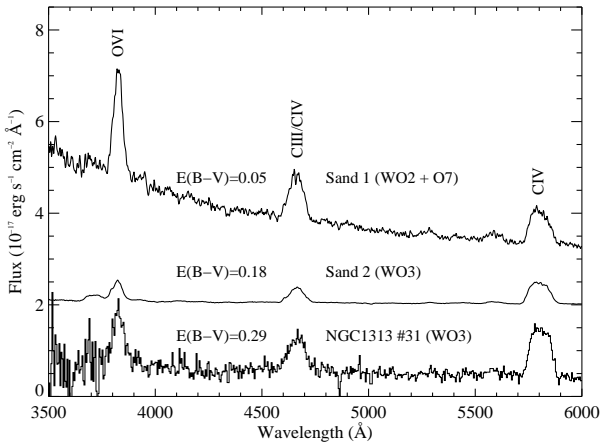,width=\columnwidth,angle=90.}}
\caption{Dereddened spectral comparison between NGC\,1313~\#31 and
  Sand\,1 (WO2 + O7) in the SMC and Sand\,2 (WO3) in the LMC.  The
  spectra for Sand\,1 and 2 are taken from \citet{kingsburgh95} and
  have been scaled to the distance of NGC\,1313 and extinction
  corrected by 0.05 and 0.18\,mag, respectively.  For clarity,
  individual spectra have been offset by 2$\times
  10^{-18}$erg\,s$^{-1}$cm$^{-1}$\AA$^{-1}$. }
\label{wo}
\end{figure}

In Fig.~\ref{wo} we compare the spectrum of NGC\,1313~\#31 with LMC
and SMC WO counterparts Sand\,1 and 2 from \citet{kingsburgh95},
clearly illustrating the presence of broad O\,{\sc vi}
$\lambda$3811--34, C\,{\sc iv} $\lambda$4658 and C\,{\sc iv}
$\lambda$5808 emission features.  To our knowledge this represents the
first detection of a WO star beyond the Local Group. Large line widths
(FWHM($\lambda$5808)$\sim$100\AA) associated with \#31 are apparent, and
are observed to be comparable with those in Local Group WO counterparts.

For spectral classification we refer to the scheme of \citet{pac98},
where the ratio of the equivalent widths (W$_{\lambda}$) of O\,{\sc
vi} $\lambda$3811--34 and O\,{\sc v} $\lambda$5590 is
used as the primary subtype discriminator.  In our data, O\,{\sc v}
$\lambda$5590 is extremely weak and accurately classifying it
based on this criterion is not possible.  However, a classification can
be made using the secondary line diagnostics O\,{\sc vi}
$\lambda$3811--34 and C\,{\sc iv} $\lambda$5808, which
also suggests a WO3 classification.

%\begin{figure*}
%\begin{tabular}{cc}
%\psfig{figure=wo_fors1_new.eps,width=\columnwidth,angle=0.}&
%\psfig{figure=wo_acs_new.eps,width=\columnwidth,angle=0.}
%\end{tabular}
%\caption{6 $\times$ 6-arcsec$^{2}$ VLT/FORS1 $\lambda$4684 (left) and
% {\it HST}/ACS F555W (right) image of NGC\,1313~\#31.  \#31 has
% been identified and marked on both images. North is up and East is to
% the left.}
%\label{wo_acs}
%\end{figure*}

%Fig~\ref{wo_acs} compares a 0.8\arcsec\ VLT/FORS1 image of \#31 with
%that acquired with HST/ACS F555W.   The superior spatial resolution of
%{\it HST}/ACS is immediately apparent and if NGC\,1313~\#31 were to
%explode as a supernova it is clear how valuable image spatial
%resolution would be for identifying possible progenitors.  

NGC\,1313~\#31 is below the detection limit of our FORS1 B-band photometry,
suggesting m$_{B}$$\geq$23.5\,mag.  {\it HST}/ACS photometry of \#31 yields
m$_{F435W}$=23.4$\pm$0.01\,mag, m$_{F555W}$=23.8$\pm$0.01\,mag,
m$_{F814W}$=25.4$\pm$0.05\,mag (Pellerin, private communication).  
%
%Assuming that WO subtypes share similar intrinsic and spectral
%properties, 

Single/binary WO stars are observed to exhibit M$_{V}\sim-3$ and
--6\,mag, respectively \citep{kingsburgh95}.  For a total extinction
E$_{B-V}$=0.29\,mag, we find M$_{V}\sim$--5\,mag for \#31, which is
very bright for a single WO star, so \#31 likely belongs to a binary
system or small association.

\section{The Global WR content of NGC\,1313}
\label{discussion}

Of the 94 WR candidates identified from VLT/FORS1 narrow-band imaging
we have identified characteristic WR signatures in 70 regions from our
spectroscopy to date. Using LMC template WR stars as local analogues,
we have derived a global WR population of 84 stars, as summarised in
Table~\ref{WR:pop}.

\subsection{Nature of Candidates}

In general, are the confirmed WR sources in NGC\,1313 isolated WR
stars, binaries  or members of luminous stellar clusters/associations?  One
may expect the latter since our 0\arcsec.8 slit width corresponds to
a spatial scale of 15\,pc at a distance of 4.1\,Mpc.  

It is well known that WC stars exhibit a larger
$\lambda$4684 excess than their WN counterparts due to their
intrinsically stronger emission features.  Synthetic filter photometry
of LMC template stars from \citet{pac06} suggests that a typical WC4
star in NGC\,1313 should have $m_{B}$$\sim$23\,mag and $\Delta m =
m_{\lambda4684} - m_{B} \sim--2$\,mag.
For WN stars, synthetic photometry suggests that single early, mid and
late WN stars should exhibit photometric excesses of $\Delta m$=--1.2,
--0.6 and --0.4\,mag, respectively.  Single WN2--4 stars within
NGC\,1313 would have an apparent B-band magnitude of $\sim$23\,mag,
whereas late-type WN stars should be slightly brighter with
m$_{B}\sim$22\,mag.

\begin{figure}
\psfig{figure=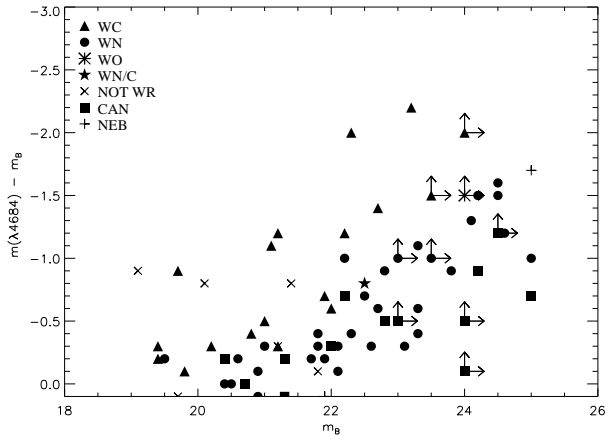,width=\columnwidth,angle=90.}
\caption{Comparison between $\Delta m \, (m_{B} - m_{\lambda4686}$)
  and m$_{B}$ for regions within NGC\,1313.  Regions which have been
  spectroscopically observed and subsequently classified or eliminated
  as WR regions are presented in the key. Also marked are candidates
  regions which still await spectroscopic follow-up.}
\label{photometry}
\end{figure}

In Fig.~\ref{photometry} we compare the actual $\lambda$4684 excess,
$\Delta m$, with the B-band magnitude for our sample.  As
expected, the fainter candidates exhibit the largest excess ($\Delta
m$$<$--1.0) consistent with single or binary systems, where the WR
emission is not strongly diluted by nearby stars.  Sources with small
photometric excess ($\Delta m$$\lesssim$--0.5) correspond to the
brighter end of our sample and are consistent with luminous
clusters/associations, where the continua of nearby stars greatly
dilute the WR emission.

Of course, intrinsic absolute magnitudes of WR stars are seen to
display a large scatter \citep{pac06}, but based on
Fig.~\ref{photometry} we conclude that three WCE sources
(NGC\,1313~\#12, 41 and 81) within NGC\,1313 are consistent with
fairly isolated WC stars.  The remaining 25 therefore belong to
binary systems or lie within stellar clusters/associations.
For the 50 WN sources, six WN2--4 and three WN7--9 stars have
properties consistent with single, isolated WN stars.

For two sources known to host WR stars we find $\Delta m>$0.5.
Since m$_{B}$$\sim$19\,mag for these regions, both are consistent with
luminous clusters/associations such that m$_{B}$ is being contaminated
by a population of stars that are intrinsically faint at
$\lambda$4684.
For the six candidates which showed no evidence for WR features we
find that the three confirmed late-type objects display significant
photometric excesses ($\Delta m \geq$--0.7\,mag).  Such sources could
be excluded from future spectroscopic surveys on the basis of their
extremely red colours, B--V$\sim$1\,mag compared with
B--V$\sim$0.1\,mag for genuine candidates.  The three other non WR
detections only display negligible photometric excesses.

Of the 16 remaining candidate regions, eight display photometric
excesses ($\Delta m \geq$--0.3\,mag) consistent with confirmed WN
sources. Indeed, three of these correspond to candidates for which WC
stars were excluded since their spectra started longward of
$\lambda$5000.  
In the continuum subtracted $\lambda$4684 image (obtained by
subtracting the $\lambda$4781 frame from the $\lambda 4684$
image), the majority of the remaining candidates are bright and are
comparable with genuine WN sources.  Overall, we expect that 11 of the
16 remaining candidates will host WR stars, some of which appear
multiple.

%Of the 12 candidate regions which were not spectroscopically observed,
%eight display photometric excesses $\Delta m$$\geq$--0.3\,mag, all of
%which are consistent with confirmed WN sources.  Indeed, three of
%these correspond to candidates for which WC stars were excluded since
%their spectra started longward of $\lambda$5000\AA.  Excesses for the
%remaining candidates are observed to be negligible, but are
%nevertheless comparable with WN regions. 

\subsection{Completeness}

\begin{figure}
\begin{tabular}{c}
\psfig{figure=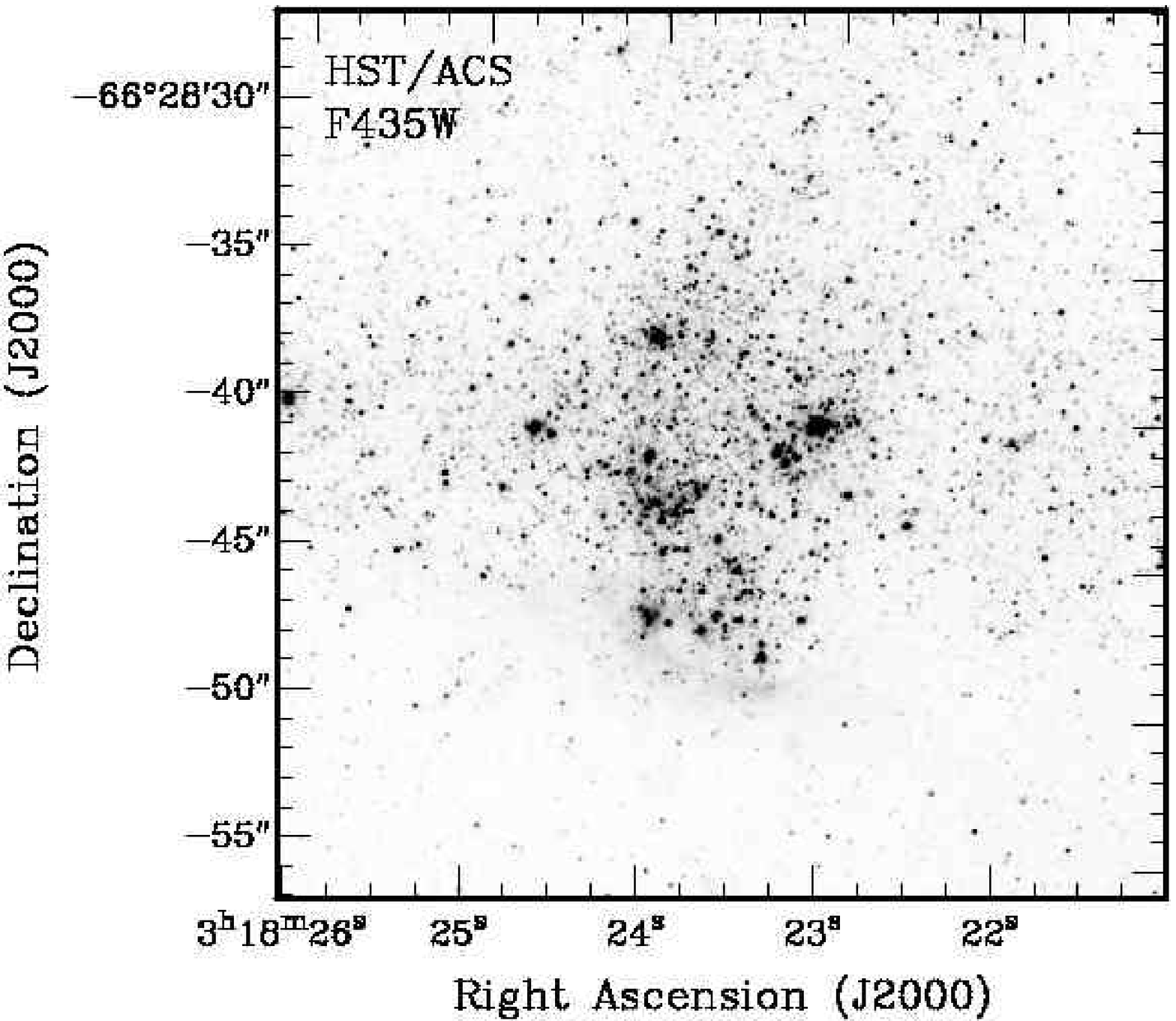,width=0.9\columnwidth,angle=0.}\\
\psfig{figure=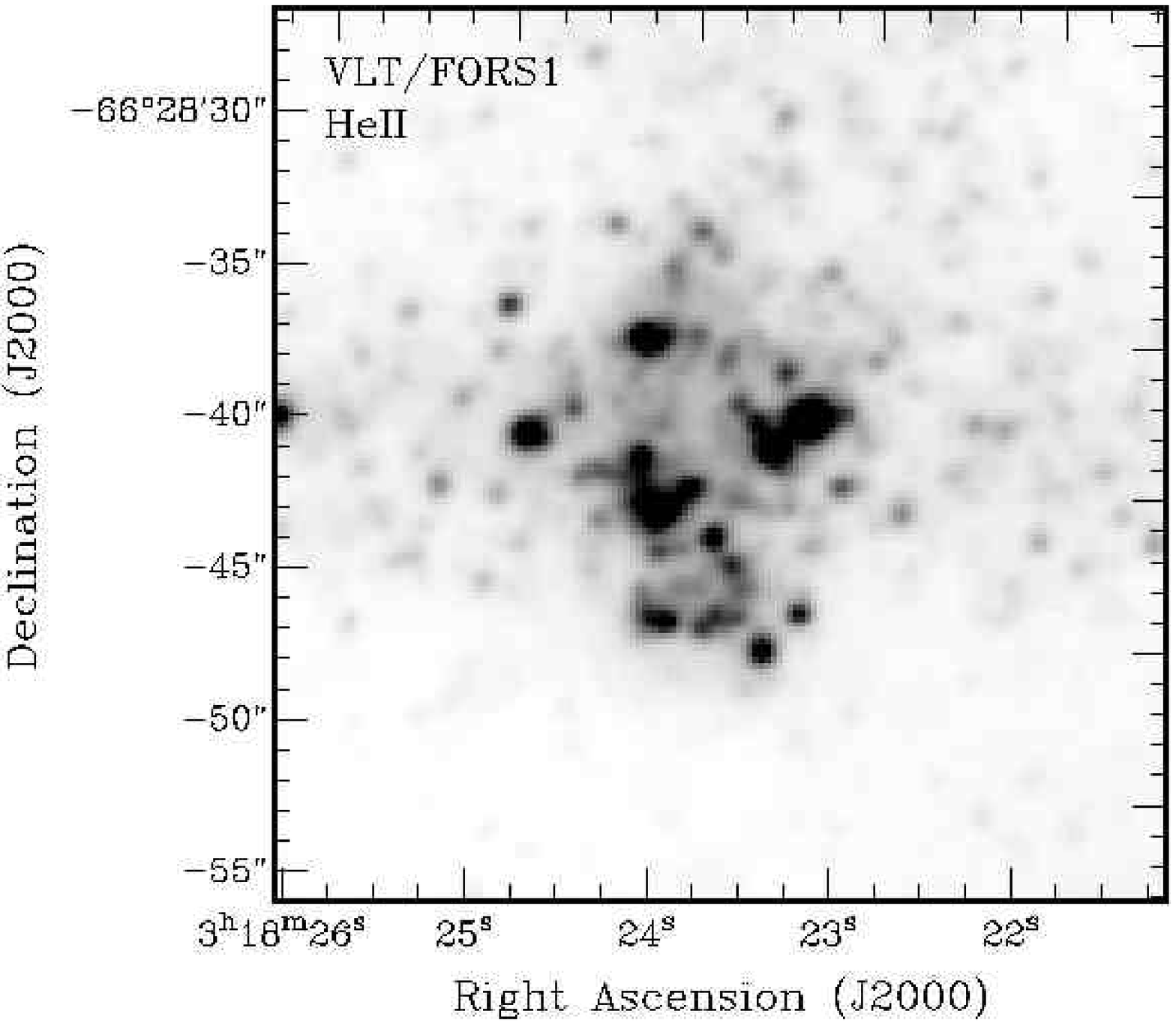,width=0.9\columnwidth,angle=0.}\\
\psfig{figure=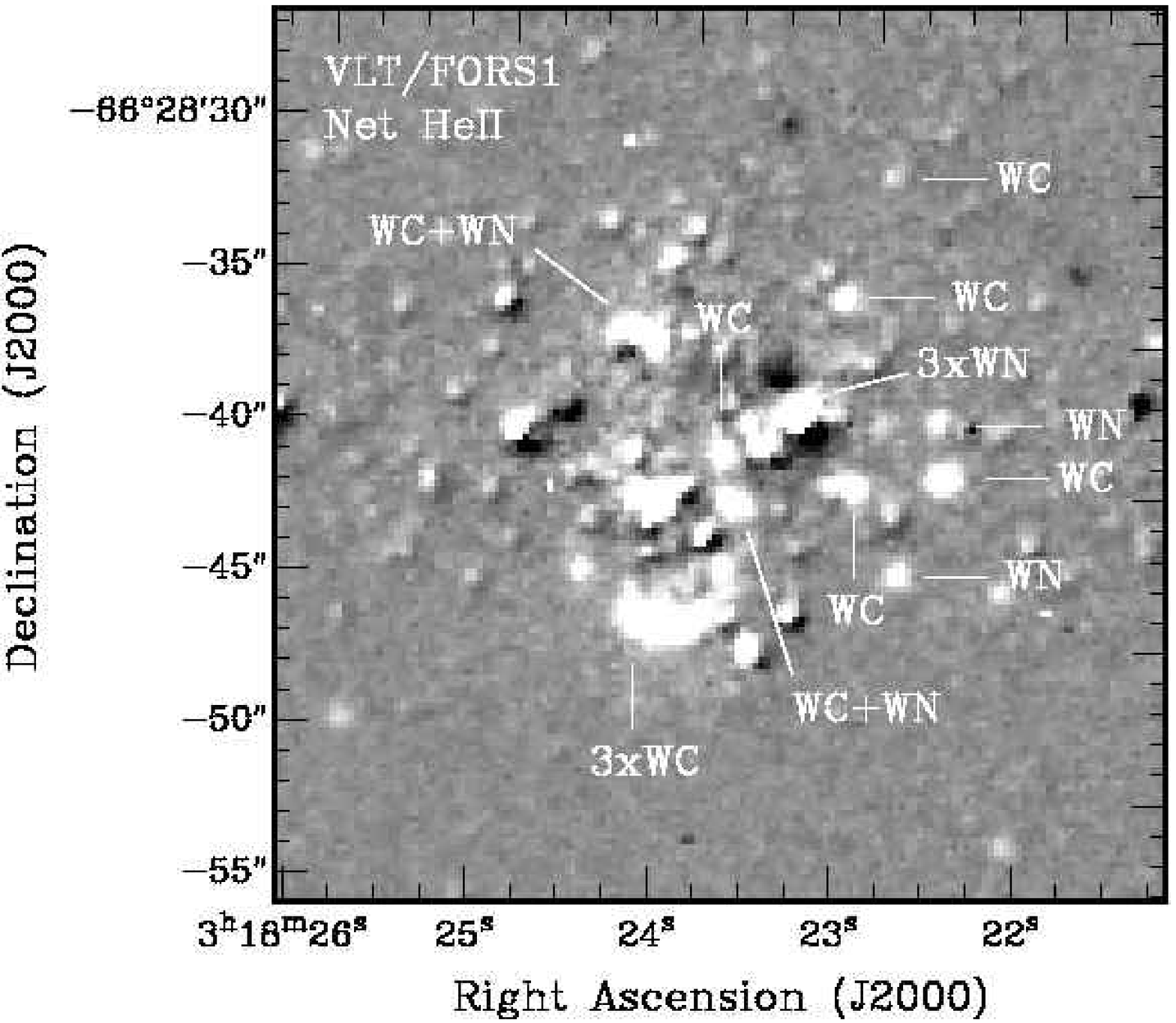,width=0.9\columnwidth,angle=0.}\\
\end{tabular}
\caption{30 $\times$ 30 \arcsec {\it HST}/ACS and VLT/FORS1
  images of the bright star forming region PES\,1.  At a distance of
  4.1\,Mpc the physical region illustrated equates to 600$\times$
  600\,pc.  The $\lambda$4684 excess sources, shown in white, which
  have been spectroscopically confirmed to host WR stars are
  marked. North is up and East is to the left on all images.
  %Top: {\it HST}/ACS F435W image of PES\,1. middle: High
  %contrast He\,{\sc ii} narrow-band FORS1 image.  %Bottom:  Difference
  %between the $\lambda$4684 and $\lambda$4781 filters, showing the
  %location of He\,{\sc ii} $\lambda$4686/C\,{\sc iii} $\lambda$4650
  %emission.  WR candidates within the region are indicated.}
  }
\label{completeness}
\end{figure}

How complete is our WR survey of NGC\,1313?  In
Fig.~\ref{completeness} we compare a FORS1 $\lambda$4684 narrow-band
image with a continuum subtracted $\lambda$4684 image of
the brightest star forming region PES\,1 (recall Table~\ref{halpha})
which highlights the complexities involved in identifying WR
candidates within crowded regions \citep[see also][]{drissen93}.

Relatively isolated WR stars are readily identified but within the
inner regions of PES\,1 it is clear that some $\lambda$4686 emission
is spatially extended on our ground-based FORS1 images.  The strong,
bright emission associated with NGC\,1313~\#64, to the south of the
H\,{\sc ii} regions, represents one of the most extreme examples.
\#64 is believed to host multiple WR stars based on the observed
C\,{\sc iv} $\lambda$5808 line luminosity
(Sect.~\ref{wcstars}). Spectroscopic observations were made with the
slit aligned perpendicular to the extended emission, consequently
$\sim$50\% of the total flux may have been missed. %The emission extends
%$\sim$3\arcsec (60\,pc at a distance of NGC\,1313) in an East-West
%direction and it is possible that only 50\% of the total flux was
%detected, 
The WR content of this region could therefore be in excess of six WR
stars.  Similar conclusions are reached for \#65, for which images
suggest $\sim$3 WR stars, but spectroscopy of \#65 is not yet
available.

Fig.~\ref{completeness} also shows a number of faint emission regions
within PES\,1, which may be genuine WR stars.  In total, spectroscopic
observations of PES\,1 indicate a minimum of 17 WR stars comprising 7
WN and 10 WC stars.  The total WR content may be $\sim$26,
significantly larger than the dozen WR stars seen within NGC\,595 in
M\,33 \citep{drissen93} and approaches the 41 WR stars
identified within the much larger 30\,Doradus region of the LMC
\citep{breysacher99}.

%12 additional WR candidates have been identified within gaint
%H\,{\sc ii} regions of NGC\,1313, eight of which are found within
%PES\,1.  Most candidates appear to be consistent with fainter WN-type
%stars sugging that the total WR population of NGC\,1313 is $\sim$115
%WR stars, with approximatly 20\% located within PES\,1.  Allowing for
%these additional WR stars N(WC)/N(WN) would decrease to
%N(WC)/N(WN)$\sim$0.4.

Completeness issues of ground-based surveys are further highlighted by
the {\it HST}/ACS F435W image of PES\,1. The superior spatial
resolution of {\it HST} is immediately apparent and \#64 and \#65 are
clearly resolved into multiple stellar clusters/associations.  If a
core-collapse supernova were to occur within either of these
regions, even with HST it may not be possible to directly identify a
WR progenitor.

Beyond PES\,1, seven other candidate $\lambda$4684 emission regions appear
spatially extended, even though observed line luminosities are consistent
with a single star in each case. These sources have been indicated in
Table~A.2.  Allowing for the possibility that a few further
$\lambda$4684 sources are WR stars (not listed in Table~A.2),
we suggest that the total WR population of NGC\,1313 is $\sim$115,
20\% of which are located within PES\,1.  Since most of the remaining
candidates are consistent with WN stars, we infer N(WC)/N(WN)=0.4.

\subsection{The WR population}

\begin{table}
\caption{WR and O star populations of NGC\,1313 and several metal-poor (log
  (O/H)+12$<$8.5) Local Group galaxies. }
\begin{tabular}{@{\hspace{1mm}}l@{\hspace{1mm}}c@{\hspace{1mm}}r@{\hspace{1mm}}r@{\hspace{1mm}}c@{\hspace{-4mm}}c@{\hspace{1mm}}c}
\hline Galaxy & log(O/H)& N(WR) & N(WR)/ &
N(WC)/&\multicolumn{1}{c}{SFR}& Ref\\
&\multicolumn{1}{c}{+12}&&\multicolumn{1}{c}{N(O)}&\multicolumn{1}{c}{N(WN)}&\\ 
& & &(10$^{-2}$)&&(M$_{\odot}\mbox{yr}^{-1}$) \\
\hline 
LMC & 8.37 & 134 &3.0&0.2&0.22&1, 2\\
IC\,10 & 8.26 &$\geq$26&$\geq$0.6&1.3&0.20 &3, 5 \\ 
NGC\,1313& 8.23 &$\sim$115 & $\sim$0.9 &$\sim$0.4&0.64&6\\
SMC & 8.13 & 12 & 1.5&0.1&0.04&1, 4\\ 
%IC\,1613 & 7.85 & \\
\hline
\end{tabular}
{\footnotesize References.---(1)\,\citet{kennicutt95};
  (2)\,\citet{breysacher99}; (3)\,\citet{pac03}; (4)\,\citet{massey03};
  (5)\,\citet{leroy06}; (6)\,this work.}\\
\label{wr:pop}
\end{table}

In Table~\ref{wr:pop} we compare the WR and O star content of
NGC\,1313 to several other Local Group, metal-poor (log(O/H)+12$<$8.5),
galaxies.
The WR content of NGC\,1313 is similar to that observed in the LMC,
yet its SFR is a factor of three higher.  Consequently, the modest
reduction in metallicity of NGC\,1313 relative to the LMC either
causes fewer O stars to advance to the WR phase, or results in shorter
WR lifetimes.  The WR content of NGC\,1313 is more comparable to that
of IC\,10, at a similar metallicity, once their relative SFR are taken
into account (see Table~\ref{wr:pop}).

%We obtain N(WR)/N(O)=115/1300$\sim$0.009, which is comparable to
%N(WR)/N(WO)$>$0.007 observed in IC\,10.

%\subsection{The WR population}

Over the last few decades WR populations have been well sampled in a
wide variety of environments and it is well known that the
distribution of WC to WN stars is dependant on metallicity
\citep{massey03}.
Large WC/WN ratios are observed in extremely metal rich environments,
e.g. N(WC)/N(WN)$\sim$1 in M\,83 \citep{hadfield05}, while small
ratios are observed in metal-poor environments e.g. N(WC)/N(WN)=0.2 in
the LMC \citep{breysacher99}.  Based on the low metallicity of
NGC\,1313 one would expect N(WC)/N(WN)$\sim$0.1.  Here, we estimate a
significantly higher subtype ratio of $\sim$0.6, (or $\sim$0.4 if we
adopt a WN subtype for the majority of our remaining candidates).  This is
intermediate between that observed in the outer, sub-solar regions of
M\,33 where N(WC)/N(WN)$\sim$0.35 \citep{massey98} and the inner
region of NGC\,300 for which N(WC)/N(WN)$\sim$0.7 \citep{pac07}.

%Overall, spiral galaxies appear to exhibit a larger WC/WN ratio
%than their irregular counterparts.  Could it be that WC stars are
%preferentially located in spiral galaxies? Of course, with only small number
%statistics it is not possible to comment here, but it
%does highlight the importance of investigating WR populations in
%different environments.

\subsection{Using the WC content as a proxy for metallicity}
 
It has long been recognised that late-type WC stars are preferentially
found within metal-rich environments.  For example, in the Milky Way
WC9 stars are universally located within the inner, metal-rich regions
\citep{conti90} and an overwhelming WC8--9 population is seen in M\,83
\citep{hadfield05}.  Conversely, metal-poor WC stars in the SMC and
LMC are predominantly early-types \citep{breysacher99}. \citet{pac02}
attributed the dominance of early WC subtypes in low-metallicity
environments to weak WC winds, since the C\,{\sc iii} $\lambda$5696
classification line scales very sensitively with wind density.  This
is further supported by the fact that the WC population of NGC\,1313
is comprised solely of early-type WC and WO stars. %and their high
%temperature WO counterparts.

\begin{figure}
   \centerline{\psfig{figure=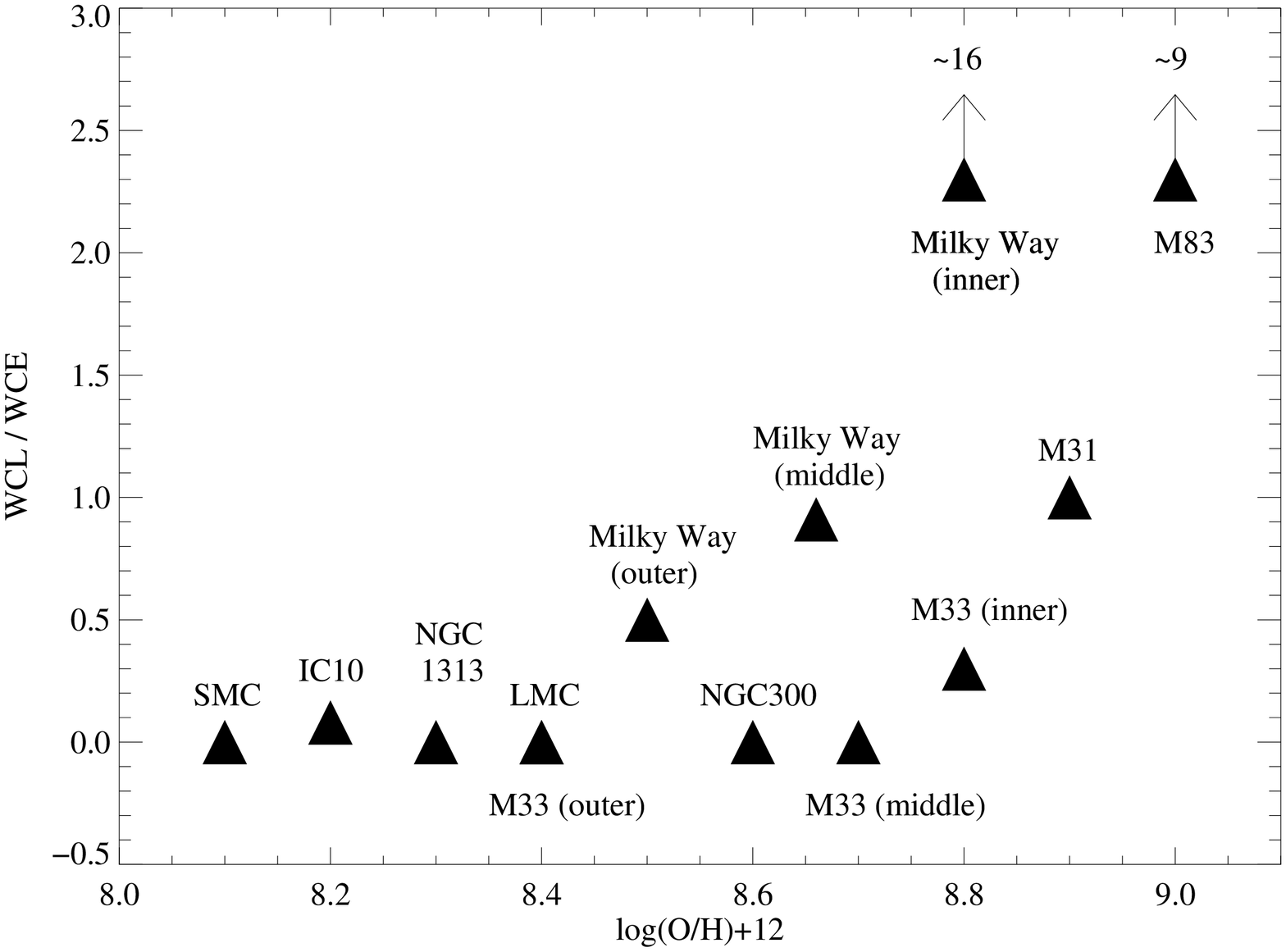,clip=,width=\columnwidth,angle=90.}}
\caption{The distribution of late (WC7--9) to early (WC4--6) carbon sequence WR stars in NGC\,1313 and
  other well studied galaxies versus metal content.  Data is taken from
  \citet{massey98}, \citet{breysacher99}, \citet{massey03},
  \citet{pac03}, \citet{abbott04}, \citet{hadfield07}, \citet{pac07}
  and the present study.}
\label{fig:wcl}
\end{figure}

In Fig.~\ref{fig:wcl} we show the subtype ratio of WC7--9 to WC4--6
stars in NGC\,1313 and other nearby galaxies.  At high metallicities
(log(O/H)+12$\geq$8.8), wind densities are sufficiently high to reveal
strong C\,{\sc iii} $\lambda$5696 emission in the majority of WC
stars, such that late WC subtypes dominate the population (e.g M\,83).
For intermediate metallicities (log(O/H)+12$\sim$8.5--8.8), the WC
population is composed of a mixture of early and late WC subtypes, as
observed for the solar neighbourhood.  For metal-poor galaxies
(log(O/H)+12$\lesssim$8.5) wind densities are low, and are
insufficient to produce late-type WC stars.  Consequently, galaxies
such as the SMC and NGC\,1313 only host early-type WC and their high
temperature WO counterparts.

Exceptions to this general trend do occur.  In the metal-poor galaxy
IC\,10 (log(O/H)+12=8.26) \citet{pac03} identify one of the 14 WC
stars as a WC7 subtype  %With this exception, it appears
%that it is possible use the dominant WC subtype as a proxy for
%metallicity.  
At high metallicities a small fraction of early WC
subtypes are also seen (e.g. M\,83).  Nevertheless, it appears that
the dominant WC subtype may serve as a crude metallicity diagnostic
for integrated stellar populations e.g. late WC subtypes for
NGC\,1365 at high-metallcity \citep{phillips92} and early-WC subtype
for NGC\,3125 at low metallicity \citep{hadfield06}.

\subsection{The origin of nebula He\,{\sc ii} $\lambda$4686}

The rarity of H\,{\sc ii} regions which show nebular He\,{\sc ii}
$\lambda$4686 among Local Group galaxies \citep{garnett91} illustrates
the significance of such regions within NGC\,1313.

The association of nebular He\,{\sc ii} $\lambda$4686 emission with
WNE and WO stars stars (e.g. NGC\,1313~\#19) in low metallicity
environments argues in favour of WR stars being the source of the high
ionization.  It has long been predicted that the hard ionizing flux
output of a WR star is dependant on wind-density \citep{schmutz92}.
Model atmospheres of \citet{smith02} predict that low wind densities
favour transparent winds such that only weak winds are expected to
produce a significant He$^{+}$ continua below 228\AA.  Consequently,
dense winds associated with most WC and WN stars are not expected to
produce strong He$^{+}$ ionizing radiation.

Since mass-loss rates of WR stars are predicted \citep{vink05}, and
observed to scale with metallicity \citep{pac02}, it is expected that
He\,{\sc ii} nebulae are preferentially associated with WR stars in
metal-poor galaxies.  \citet{smith02} predict that for WNE stars the
He$^{+}$ continua increases from log Q$_{2}$ = 41.2 at {\it Z}={\it
Z}$_{\odot}$ to 48.3 for {\it Z}=0.2{\it Z}$_{\odot}$.  In metal-rich
environments WR stars would not be capable
of ionising He$^{+}$.  Indeed, five of the six WR stars which exhibit
nebular He\,{\sc ii} are in galaxies with log(O/H)+12$<$8.4
(Table~\ref{table:nebheii}), with the association increasing from less
than 0.3\% in the Milky Way \citep[e.g. Sand 4,][]{Polcaro95} to 100\%
in IC\,1613 which is a factor of ten lower in metallicity
\citep{kingsburgh95b}.
 
\section{Summary \& Conclusions}
\label{conclusion}

We present the results of an imaging and spectroscopic survey of the
WR content of the nearby galaxy NGC\,1313.  From narrow-band
$\lambda$4684 and $\lambda$4781 images we have identified 94 candidate
He\,{\sc ii} emission regions of which 82 have been spectroscopically
observed.  Broad WR emission has been identified within 70 regions,
with only six candidates failing to show stellar/nebular He\,{\sc ii}
$\lambda$4686 emission.

A nebular analysis of several H\,{\sc ii} regions within the galaxy
confirms that NGC\,1313 has a metal content
(log(O/H)+12=8.23$\pm$0.06) intermediate between the SMC and LMC and
equivalent to that of IC\,10.  Using template LMC WR stars, we
estimate N(WR)=84, with N(WC)/(WN)$\sim$0.6 from spectroscopy.  The 51
WN stars are evenly distributed amongst early and late subtypes and
include a rare WN/C4 transition star.  The 33 WC stars are exclusively
early type stars, one of which we assign a WO classification. Indeed,
NGC\,1313~\#31 represents the first WO star identified beyond the
Local Group.

Photometry of the remaining candidates, plus a number of additional
sources identified from spatially extended $\lambda4684$ emission
(primarily within PES\,1), are mostly consistent with WN subtypes.  We
suggest a total WR content of N(WR)$\sim$115 for NGC\,1313; a factor
of four times higher than that observed in IC\,10 for which the SFR is
three times lower.  The bright star forming region PES\,1 hosts 20\%
of the entire WR content of NGC\,1313, higher than that observed in
NGC 595 in M\,33 despite a comparable H$\alpha$ luminosity.  Globally
N(WC)/N(WN)$\sim$0.4, substantially higher than that of the LMC for
which N(WC)/N(WN)=0.2.

Late-type WC stars are notably absent in NGC\,1313, a common feature
of metal-poor galaxies within the Local Group.  Indeed, one may infer
the metallicity of a spatially unresolved WR galaxy based on the
dominant WC subtype.  If the spectral appearance is dominated by late-type
WC stars then one would infer log(O/H)+12$\gtrsim$8.8.  Conversely, a
dominant early-type WC appearance suggests log(O/H)+12$\lesssim$8.5.

In addition, we have identified strong nebular He\,{\sc ii}
$\lambda$4686 in three regions within NGC\,1313.  For NGC\,1313~\#19,
the nebular emission is accompanied by broad He\,{\sc ii} emission
(consistent with an WNE star), only the sixth case believed to be
ionized by a WR star.  For the other two regions, one is consistent
with shock ionization, most probably a supernova remnant, whereas the
ionizing source of NGC\,1313~\#6 may possibly be a X-ray source,
similar to that inferred for N44C within the LMC.

\section*{Acknowledgements}
We would like to thank Anne Pellerin for providing HST photometry for
sources within NGC\,1313.  LJH acknowledges financial support from
PPARC/STFC.

\bibliographystyle{mn2e} \bibliography{abbrev,refs}

\begin{thebibliography}{}

\bibitem[\protect\citeauthoryear{Abbott, Crowther, Drissen, Dessart, Martin \&
  Boivin}{Abbott et~al.}{2004}]{abbott04}
Abbott J.~B.,  Crowther P.~A.,  Drissen L.,  Dessart L.,  Martin P.,    Boivin
  G.,  2004, MNRAS, 350, 552

\bibitem[\protect\citeauthoryear{Breysacher, Azzopardi \& Testor}{Breysacher
  et~al.}{1999}]{breysacher99}
Breysacher J.,  Azzopardi M.,    Testor G.,  1999, A\&AS, 137, 117

\bibitem[\protect\citeauthoryear{Colbert, Petre R. \& Ryder}{Colbert
  et~al.}{1995}]{colbert95}
Colbert E. J.~M.,  Petre R. S. E.~M.,    Ryder S.~D.,  1995, ApJ, 446, 177

\bibitem[\protect\citeauthoryear{Conti \& Vacca}{Conti \&
  Vacca}{1990}]{conti90}
Conti P.~S.,  Vacca W.~D.,  1990, AJ, 100, 2

\bibitem[\protect\citeauthoryear{Crowther}{Crowther}{2007}]{pac07b}
Crowther P.~A.,  2007, A\&AR, 45, 177

\bibitem[\protect\citeauthoryear{Crowther, Carpano, Hadfield \&
  Pollock}{Crowther et~al.}{2007}]{pac07}
Crowther P.~A.,  Carpano S.,  Hadfield L.~J.,    Pollock A. M.~T.,  2007, A\&A,
  469, L31

\bibitem[\protect\citeauthoryear{Crowther, De~{M}arco \& Barlow}{Crowther
  et~al.}{1998}]{pac98}
Crowther P.~A.,  De~{M}arco O.,    Barlow M.~J.,  1998, MNRAS, 296, 367

\bibitem[\protect\citeauthoryear{Crowther, Dessart, Hillier, Abbott \&
  Fullerton}{Crowther et~al.}{2002}]{pac02}
Crowther P.~A.,  Dessart L.,  Hillier D.~J.,  Abbott J.,    Fullerton A.~W.,
  2002, A\&A, 392, 653

\bibitem[\protect\citeauthoryear{Crowther, Drissen, Abbott, Royer \&
  Smartt}{Crowther et~al.}{2003}]{pac03}
Crowther P.~A.,  Drissen L.,  Abbott J.~B.,  Royer P.,    Smartt S.~J.,  2003,
  A\&A, 404, 483

\bibitem[\protect\citeauthoryear{Crowther \& Hadfield}{Crowther \&
  Hadfield}{2006}]{pac06}
Crowther P.~A.,  Hadfield L.~J.,  2006, A\&A, 449, 711

\bibitem[\protect\citeauthoryear{Crowther, Smith \& Willis}{Crowther
  et~al.}{1995}]{pac95}
Crowther P.~A.,  Smith L.~J.,    Willis A.~J.,  1995, A\&A, 304, 269

\bibitem[\protect\citeauthoryear{Drissen, Moffat \& Shara}{Drissen
  et~al.}{1993}]{drissen93}
Drissen L.,  Moffat A. F.~J.,    Shara M.~M.,  1993, A\&A, 105, 1400

\bibitem[\protect\citeauthoryear{Friedli, Benz \& Kennicutt}{Friedli
  et~al.}{1994}]{friedli94}
Friedli D.,  Benz W.,    Kennicutt R.,  1994, ApJ, 430, 105

\bibitem[\protect\citeauthoryear{Garnett, Galarza \& Chu}{Garnett
  et~al.}{2000}]{garnett00}
Garnett D.,  Galarza V.~C.,    Chu Y.-H.,  2000, AJ, 545, 251

\bibitem[\protect\citeauthoryear{Garnett, Kennicutt, Chu \& Skillman}{Garnett
  et~al.}{1991}]{garnett91}
Garnett D.,  Kennicutt R.~C.,  Chu Y.~H.,    Skillman E.~D.,  1991, AJ, 373,
  458

\bibitem[\protect\citeauthoryear{Hadfield \& Crowther}{Hadfield \&
  Crowther}{2006}]{hadfield06}
Hadfield L.~J.,  Crowther P.~A.,  2006, MNRAS, 368, 1822

\bibitem[\protect\citeauthoryear{Hadfield, Crowther, Schild \&
  Schmutz}{Hadfield et~al.}{2005}]{hadfield05}
Hadfield L.~J.,  Crowther P.~A.,  Schild H.,    Schmutz W.,  2005, A\&A, 439,
  265

\bibitem[\protect\citeauthoryear{Hadfield, van Dyk, Morris, Smith, Marston \&
  Peterson}{Hadfield et~al.}{2007}]{hadfield07}
Hadfield L.~J.,  van Dyk S.~D.,  Morris P.~W.,  Smith J.~D.,  Marston A.~P.,
  Peterson D.~E.,  2007, MNRAS, 376, 248

\bibitem[\protect\citeauthoryear{Hummer \& Storey}{Hummer \&
  Storey}{1987}]{hummer87}
Hummer D.,  Storey P.~J.,  1987, MNRAS, 224, 801

\bibitem[\protect\citeauthoryear{Kennicutt}{Kennicutt}{1984}]{kennicutt84}
Kennicutt R.~C.,  1984, ApJ, 287, 116

\bibitem[\protect\citeauthoryear{Kennicutt}{Kennicutt}{1998}]{kennicutt98}
Kennicutt R.~C.,  1998, ARA\&A, 36, 189

\bibitem[\protect\citeauthoryear{Kennicutt, Bresolin, Bomans, Bothun \&
  Thompson}{Kennicutt et~al.}{1995}]{kennicutt95}
Kennicutt R.~C.,  Bresolin F.,  Bomans D.~J.,  Bothun G.~D.,    Thompson I.,
  1995, AJ, 109, 594

\bibitem[\protect\citeauthoryear{Kennicutt \& Kent}{Kennicutt \&
  Kent}{1983}]{kennicutt83}
Kennicutt R.~C.,  Kent S.~M.,  1983, AJ, 88, 1094

\bibitem[\protect\citeauthoryear{Kingsburgh \& Barlow}{Kingsburgh \&
  Barlow}{1995}]{kingsburgh95b}
Kingsburgh R.,  Barlow M.~J.,  1995, A\&A, 295, 171

\bibitem[\protect\citeauthoryear{Kingsburgh, Barlow \& Storey}{Kingsburgh
  et~al.}{1995}]{kingsburgh95}
Kingsburgh R.,  Barlow M.~J.,    Storey P.,  1995, A\&A, 295, 75

\bibitem[\protect\citeauthoryear{Landolt}{Landolt}{1992}]{landolt92}
Landolt A.~U.,  1992, AJ, 104, 340

\bibitem[\protect\citeauthoryear{Larsen}{Larsen}{2004}]{Larsen04}
Larsen S.,  2004, A\&A, 416, 537

\bibitem[\protect\citeauthoryear{Leroy, Bolatto A.and~Walter \& Blitz}{Leroy
  et~al.}{2006}]{leroy06}
Leroy A.,  Bolatto A.and~Walter F.,    Blitz L.,  2006, ApJ, 643, 825

\bibitem[\protect\citeauthoryear{Marcelin \& Gondoin}{Marcelin \&
  Gondoin}{1983}]{marcelin83}
Marcelin M.,  Gondoin P.,  1983, A\&AS, 51, 3538

\bibitem[\protect\citeauthoryear{Massey \& Johnson}{Massey \&
  Johnson}{1998}]{massey98}
Massey P.,  Johnson O.,  1998, ApJ, 505, 793

\bibitem[\protect\citeauthoryear{Massey, Olsen \& Parker}{Massey
  et~al.}{2003}]{massey03}
Massey P.,  Olsen K. A.~G.,    Parker J.~W.,  2003, PASP, 115, 1265

\bibitem[\protect\citeauthoryear{M{\'e}ndez, Davis, Moustakas, Newman, Madore
  \& Freedman}{M{\'e}ndez et~al.}{2002}]{mendez02}
M{\'e}ndez B.,  Davis M.,  Moustakas J.,  Newman J.,  Madore B.~F.,    Freedman
  W.~L.,  2002, AJ, 124, 213

\bibitem[\protect\citeauthoryear{Miller, Schlegel, Petre \& Colbert}{Miller
  et~al.}{1998}]{miller98}
Miller S.,  Schlegel E.~M.,  Petre R.,    Colbert E.,  1998, AJ, 116, 1657

\bibitem[\protect\citeauthoryear{Naz{\'e}, Rauw, Manfroid, Chu \&
  Vreux}{Naz{\'e} et~al.}{2003a}]{naze03}
Naz{\'e} Y.,  Rauw G.,  Manfroid J.,  Chu Y.-U.,    Vreux J.-M.,  2003a, A\&A,
  408

\bibitem[\protect\citeauthoryear{Naz{\'e}, Rauw, Manfroid, Chu \&
  Vreux}{Naz{\'e} et~al.}{2003b}]{naze03b}
Naz{\'e} Y.,  Rauw G.,  Manfroid J.,  Chu Y.-U.,    Vreux J.-M.,  2003b, A\&A,
  L13

\bibitem[\protect\citeauthoryear{Pagel, Edmunds \& Smith}{Pagel
  et~al.}{1980}]{pagel80}
Pagel B. E.~J.,  Edmunds M.~G.,    Smith G.,  1980, MNRAS, 193, 219

\bibitem[\protect\citeauthoryear{Pellerin, Meyer, Harris \& Calzetti}{Pellerin
  et~al.}{2007}]{pellerin07}
Pellerin A.,  Meyer M.,  Harris J.,    Calzetti D.,  2007, AJ, 658, L87

\bibitem[\protect\citeauthoryear{Phillips \& Conti}{Phillips \&
  Conti}{1992}]{phillips92}
Phillips A.,  Conti P.,  1992, ApJ, 385, L91

\bibitem[\protect\citeauthoryear{Polcaro, Rossi, Norci \& Viotti}{Polcaro
  et~al.}{1995}]{Polcaro95}
Polcaro V.~F.,  Rossi C.,  Norci L.,    Viotti R.,  1995, A\&A, 303, 211

\bibitem[\protect\citeauthoryear{Russell \& Dopita}{Russell \&
  Dopita}{1990}]{russell90}
Russell S.~C.,  Dopita M.,  1990, ApJS, 74, 93

\bibitem[\protect\citeauthoryear{Ryder \& Dopita}{Ryder \&
  Dopita}{1994}]{ryder94}
Ryder S.~D.,  Dopita M.~A.,  1994, ApJ, 430, 142

\bibitem[\protect\citeauthoryear{Schaerer, Contini \& Pindao}{Schaerer
  et~al.}{1999}]{schaerer99}
Schaerer D.,  Contini T.,    Pindao M.,  1999, A\&A, 341, 399

\bibitem[\protect\citeauthoryear{Schaerer \& Vacca}{Schaerer \&
  Vacca}{1998}]{sv98}
Schaerer D.,  Vacca W.~D.,  1998, ApJ, 497, 618

\bibitem[\protect\citeauthoryear{Schild, Crowther, Abbott \& Shmutz}{Schild
  et~al.}{2003}]{schild03}
Schild H.,  Crowther P.~A.,  Abbott J.~B.,    Shmutz W.,  2003, A\&A, 397, 859

\bibitem[\protect\citeauthoryear{Schlegel, Finkbeiner \& Davis}{Schlegel
  et~al.}{1998}]{schlegel98}
Schlegel D.~J.,  Finkbeiner D.~P.,    Davis M.,  1998, ApJ, 500, 525

\bibitem[\protect\citeauthoryear{Schmutz, Leitherer \& Gruenwald}{Schmutz
  et~al.}{1992}]{schmutz92}
Schmutz W.,  Leitherer C.,    Gruenwald R.,  1992, PASP, 104, 1164

\bibitem[\protect\citeauthoryear{Seaton}{Seaton}{1979}]{seaton79}
Seaton M.~J.,  1979, MNRAS, 187, 73{\sc p}

\bibitem[\protect\citeauthoryear{Smith, Norris \& Crowther}{Smith
  et~al.}{2002}]{smith02}
Smith L.~F.,  Norris R. F.~P.,    Crowther P.~A.,  2002, MNRAS, 337, 1309

\bibitem[\protect\citeauthoryear{Smith, Shara \& Moffat}{Smith
  et~al.}{1990}]{smith90}
Smith L.~F.,  Shara M.~M.,    Moffat A.~J.,  1990, ApJ, 348, 471

\bibitem[\protect\citeauthoryear{Smith, Shara \& Moffat}{Smith
  et~al.}{1996}]{smith96}
Smith L.~F.,  Shara M.~M.,    Moffat A.~J.,  1996, MNRAS, 281, 229

\bibitem[\protect\citeauthoryear{Smith, Kirshner, Blair, Long \& Winkler}{Smith
  et~al.}{1993}]{smith93}
Smith R.~C.,  Kirshner R.~P.,  Blair W.~B.,  Long K.~S.,    Winkler P.~F.,
  1993, ApJ, 407, 564

\bibitem[\protect\citeauthoryear{van~der Hucht}{van~der
  Hucht}{2006}]{derhucht06}
van~der Hucht K.~A.,  2006, A\&A, 458, 453

\bibitem[\protect\citeauthoryear{Vink \& de {K}oter}{Vink \&
  de~{K}oter}{2005}]{vink05}
Vink J.~S.,  de {K}oter A.,  2005, A\&A, 442, 587

\bibitem[\protect\citeauthoryear{Walsh \& Roy}{Walsh \& Roy}{1997}]{walsh97}
Walsh J.~R.,  Roy J.~R.,  1997, MNRAS, 288, 726

\end{thebibliography}

\vfill
\pagebreak

\appendix

\begin{table}
{Table A.1. Catalogue of H\,{\sc ii} regions observed with NGC\,1313.}%{tab:hii}}
\begin{tabular}{rrrccc}
\hline
 &\multicolumn{1}{c}{RA} &
 \multicolumn{1}{c}{Dec}&E$(B-V)$&\multicolumn{2}{c}{H\,{\sc ii} Association}\\
\multicolumn{1}{c}{\#}&\multicolumn{2}{c}{(J2000)}& &PES & W\&R\\
\hline
    H1  & 03:17:57.2 & -66:29:43.6 & 0.22 & --& --  \\% H7  1445 1098
    H2  & 03:17:57.3 & -66:31:39.5 & 0.20 & --& --  \\% H11 1294  540
    H3  & 03:18:05.3 & -66:30:27.8 & 0.19 & 6 & 20  \\% C4  1156  947
    H4  & 03:18:05.3 & -66:30:53.7 & 0.28 & --& --  \\% H6  1123  822
    H5  & 03:18:07.8 & -66:31:21.2 & 0.26 &   & 23  \\% C1  1015  709
    H6  & 03:18:24.8 & -66:31:46.3 & 0.26 & --& --  \\% H2   494  718
    H7  & 03:18:25.9 & -66:27:43.0 & 0.35 & --& --  \\% H5   772 1900
    H8  & 03:18:33.3 & -66:27:42.1 & 0.56 & --& --  \\% H3   559 1961
    H9  & 03:18:40.9 & -66:29:30.5 & 0.24 & 10 &  3  \\% H1   203 1496
\hline
\end{tabular}
\end{table}

\begin{landscape}
\begin{table}
{Table A.2. Catalogue of candidate He\,{\sc ii} emission regions
within NGC\,1313. Association with H\,{\sc ii} regions identified by
\citet[W\& R,][]{walsh97} and \citet[PES,][]{pagel80} are also given.
Note that values in parenthesis correspond to candidates which reside
close to W\&R and PES regions.  We present line properties and
spectral types for those regions which have been classified via
follow-up spectroscopy.  Sources believed to host multiple WRs based
on the observed line flux or appearance on the continuum subtracted
$\lambda$4684 image have been identified.  Regions which are awaiting
follow-up spectroscopy have been labelled WN? or NON-WR? depending on
their photometric excess and appearance on the continuum subtracted
$\lambda$4684 image. De-projected distances are expressed as a
fraction of the Holmberg radius $\rho_{0} = 4.6\arcmin =
5.5\,\mbox{kpc}$.Dereddened line fluxes have been derived assuming
E$_{B-V}$=0.29\,mag and are expressed in $\mbox{erg s}^{-1}
\mbox{cm}^{-2}$. The number of WR stars in each source is estimated
using the line luminosity calibrations of \citet*{pac06} and a
distance modulus of 28.08\,mag \citep{mendez02}.}\\
% \label{tab:candidates}}
\begin{tabular}{rrrcccrrrrrrrrl}
\hline \multicolumn{1}{c}{\#}&\multicolumn{1}{c}{$\alpha$} &
\multicolumn{1}{c}{$\delta$}&$\rho/\rho_{0}$& \multicolumn{2}{c}{H\,{\sc ii} Assoc.}& 
m$_{B}$&m$_{4684}$& \multicolumn{1}{c}{$\Delta m$} & F($\lambda$4650)+ &   L($\lambda$4650)+ &  F($\lambda$5808) &L($\lambda$5808)& Sp & Remarks\\
&\multicolumn{2}{c}{(J2000)}&&W\&R&PES&(mag)&(mag)&(mag)&F($\lambda$4686)&L($\lambda$4686)&&&Type\\
\hline 
 1 & 03:17:40.3 & -66:31:30.4 & 1.25 &  	&     &   22.5 &   21.9 &  --0.7       & \multicolumn{4}{c}{No data available}							 &  WNE     &  $\geq$1WN  \\% 95	
 2 & 03:17:41.0 & -66:31:37.1 & 1.24 &  	&     &   22.7 &   21.2 &  --1.4       & \multicolumn{2}{c}{No blue data}  			                 & 1.1$\times 10^{-15}$  &4.9$\times 10^{36}$&  WC4--5	    \\% 94   
 3 & 03:17:43.6 & -66:31:15.2 & 1.13 &  	&     &   22.1 &   21.8 &  --0.3       & 1.6$\times 10^{-16}$	 &   9.1$\times 10^{35}$ & 			 &      	     &  WN7--9  	    \\% 92   
 4 & 03:17:46.0 & -66:31:47.2 & 1.09 &          &     &   21.3 &   21.1	&  --0.2       &\multicolumn{2}{c}{No blue data}                 &         	         &                                    &            & NON-WR? \\ % 90   
 5 & 03:17:47.7 & -66:29:45.8 & 0.94 &  	&     &   21.8 &   21.8 &    0.0       & 			 &  			 & 			 &      	     &  	   & NON-WR   \\% 91  
 6 & 03:17:47.7 & -66:29:37.4 & 0.95 &  	&     &   25.0 &   23.3 &  --1.7       & 			 &  			 & 			 &      	     &  	   &Neb He\,{\sc ii}	    \\% 93   
 7 & 03:17:57.8 & -66:33:05.1 & 0.93 & (27)	&     &   20.4 &   20.4 &    0.0       & 2.6$\times 10^{-16}$	 &   1.5$\times 10^{36}$ & 			 &      	     &  WN7--9  	    \\% 80   
 8 & 03:17:58.0 & -66:33:07.9 & 0.93 &  27	&     &   20.1 &   20.3 &    0.2       & 2.5$\times 10^{-16}$	 &   1.4$\times 10^{36}$ & 			 &      	     &  WN7--9  	    \\% 77   
 9 & 03:17:59.3 & -66:30:12.0 & 0.56 &  	&     &   23.1 &   22.8 &  --0.3       & 7.2$\times 10^{-17}$	 &   4.0$\times 10^{35}$ & 			 &      	     &  WN2--4      \\% 88   
\smallskip
10 & 03:17:59.7 & -66:30:06.3 & 0.54 &  22	&     &   19.4 &   19.2 &  --0.3       & 1.5$\times 10^{-15}$	 &   8.6$\times 10^{36}$ & 1.7$\times 10^{-15}$  &7.3$\times 10^{36}$&  WC4--5	 & 2$\times$WC?  \\% 87    
11 & 03:18:01.0 & -66:29:29.8 & 0.50 &  	&     &   22.5 &   21.7 &  --0.8       & 3.7$\times 10^{-16}$	 &   2.1$\times 10^{36}$ & 5.7$\times 10^{-16}$  &2.5$\times 10^{36}$&  WN5--6/C4	  \\% 89     
12 & 03:18:01.9 & -66:30:09.5 & 0.47 &  	&     &   22.3 &   20.3 &  --2.0       & 1.5$\times 10^{-15}$	 &   8.6$\times 10^{36}$ & 1.0$\times 10^{-15}$  &4.5$\times 10^{36}$&  WC4--5	 & Mulitiple  WRs?   \\% 86    
13 & 03:18:02.7 & -66:32:51.6 & 0.79 &  	&     &   22.7 &   22.2 &  --0.6       & 2.5$\times 10^{-16}$	 &   1.4$\times 10^{36}$ & 			 &      	     &  WN7--9  	    \\% 71   
14 & 03:18:02.8 & -66:30:14.9 & 0.44 &  21	& 6   &   17.1 &   17.8 &    0.8       & 4.7$\times 10^{-16}$	 &   2.7$\times 10^{36}$ & 			 &      	     &  WN5--6?  & Mulitiple  WRs?	    \\% 85   
15 & 03:18:03.7 & -66:32:32.9 & 0.71 &  25	& 7   &   20.8 &   20.5 &  --0.4       & \multicolumn{2}{c}{No blue data}  			 & 8.7$\times 10^{-16}$  &3.8$\times 10^{36}$&  WC4--5	    \\% 70   
16 & 03:18:03.7 & -66:30:23.1 & 0.42 &  	&     &   21.2 &   20.9 &  --0.3       & 9.1$\times 10^{-16}$	 &   5.1$\times 10^{36}$ & 9.0$\times 10^{-16}$  &3.8$\times 10^{36}$&  WC4--5	& Mulitiple  WRs?   \\% 82 + 83  
17 & 03:18:03.9 & -66:30:19.5 & 0.41 &  	&     &   ---  &   ---  &    ---       & 1.9$\times 10^{-15}$	 &  			 & 			 &1.1$\times 10^{37}$&  WN	& Multiple  WRs?    \\% 81a	     
18 & 03:18:04.0 & -66:30:01.8 & 0.39 &  	&     &   24.1 &   22.8 &  --1.3       & 1.2$\times 10^{-16}$	 &   6.8$\times 10^{35}$ & 			 &      	     &  WN7--9  	    \\% 84 & 
19 & 03:18:04.4 & -66:32:46.6 & 0.74 &  	&     &   23.8 &   23.0 &  --0.9       & 1.4$\times 10^{-16}$	 &   8.0$\times 10^{35}$ & 			 &      	     &  WN2--4  	    \\% 69 & 
\smallskip
20 & 03:18:04.4 & -66:30:21.2 & 0.39 &  	&     &   22.0 &   21.4 &  --0.6       & 3.4$\times 10^{-16}$	 &   1.9$\times 10^{36}$ & 3.3$\times 10^{-16}$  &1.4$\times 10^{36}$&  WC4--5	   \\% 81 & 
21 & 03:18:04.9 & -66:30:27.0 & 0.38 &  20	& 5   &   20.6 &   20.8 &    0.2       & 1.7$\times 10^{-16}$	 &   9.4$\times 10^{35}$ & 			 &      	     &  WN7--9  	    \\% 79 & 
22 & 03:18:05.2 & -66:30:26.2 & 0.37 &  20	& 5   &   17.8 &   18.8 &    1.0       & 			 &  			 & 			 &      	     &  WN7--9  	    \\% 78 & 
23 & 03:18:06.1 & -66:30:38.0 & 0.36 &  	&     &   23.3 &   22.9 &  --0.4       & 2.0$\times 10^{-16}$	 &   1.1$\times 10^{36}$ & 			 &      	     &  WN2--4  	    \\% 75 & 
24 & 03:18:06.7 & -66:30:21.5 & 0.32 &  	&     &   21.0 &   20.7 &  --0.3       & 2.4$\times 10^{-16}$	 &   1.3$\times 10^{36}$ & 			 &      	     &  WN7--9  	    \\% 76 & 
25 & 03:18:09.1 & -66:30:04.6 & 0.22 &  	&     &   24.2 &   22.7 &  --1.5       & 1.8$\times 10^{-16}$	 &   1.0$\times 10^{36}$ & 			 &      	     &  WN7--9  	    \\% 73 & 
26 & 03:18:10.9 & -66:31:07.7 & 0.32 &  	&     &   19.1 &   18.3 &  --0.9       & 			 &  			 & 			 &      	     &  	   & Late   \\% 64 & 
27 & 03:18:11.1 & -66:29:34.8 & 0.16 &  	&     &   21.2 &   21.0 &  --0.3       & 			 &  			 & 			 &      	     &  	   & NON-WR  \\% 72  
28 & 03:18:11.7 & -66:28:42.2 & 0.28 &  	&     &  $>$23 &   $>$22.5 &  $<$--0.5    & 			 &  			 & 			 &      	     &  	   & WN?    \\% 74  
29 & 03:18:12.7 & -66:30:54.2 & 0.25 &  	&     &   23.3 &   22.3 &  --1.1       & 2.0$\times 10^{-16}$	 &   1.1$\times 10^{36}$ & 			 &      	     &  WN2--4  	    \\% 60  
\smallskip
30 & 03:18:13.4 & -66:32:16.0 & 0.53 &  	&     &   21.9 &   21.7 &  --0.2       & 7.3$\times 10^{-17}$	 &   4.1$\times 10^{35}$ & 			 &      	     &  WN7--9  	    \\% 45  
31 & 03:18:13.5 & -66:29:30.9 & 0.10 &  	&     &  $>$24 &   23.3 &  $<$--1       & 2.9$\times 10^{-16}$	 &   1.7$\times 10^{36}$ & 5.7$\times 10^{-16}$  &2.5$\times 10^{36}$&  WO3	   \\% 68  
32 & 03:18:13.6 & -66:32:17.1 & 0.54 &  	&     &   22.3 &   21.9 &  --0.4       & 1.1$\times 10^{-16}$	 &   6.0$\times 10^{35}$ & 			 &      	     &  WN7--9  	    \\% 43  
33 & 03:18:13.9 & -66:30:24.1 & 0.13 &  	&     &   \multicolumn{3}{c}{Extended} & 4.7$\times 10^{-16}$	 &   2.6$\times 10^{36}$ & 			 &      	     &  WN2--4     & 3$\times$WN?      \\% 62  
34 & 03:18:14.2 & -66:29:55.9 & 0.05 &  	&     &   22.0 &   21.7 &  --0.3       & 			 &  			 & 			 &      	     &  	   & WN?      \\% 63  
35 & 03:18:14.9 & -66:30:21.5 & 0.11 &  17	&     &   19.4 &   19.2 &  --0.2       & 9.0$\times 10^{-16}$	 &   5.1$\times 10^{36}$ & 8.0$\times 10^{-16}$  &3.5$\times 10^{36}$&  WC4--5   \\% 57  
36 & 03:18:15.0 & -66:29:10.4 & 0.15 &  	&     &   20.1 &   19.3 &  --0.8       & 			 &  			 & 			 &      	     &  	   & Late \\% 67  
37 & 03:18:15.4 & -66:29:54.0 & 0.01 &  	&     &   \multicolumn{3}{c}{Extended} & 1.1$\times 10^{-15}$	 &   6.0$\times 10^{36}$ & 7.8$\times 10^{-16}$  &3.5$\times 10^{36}$&  WC4--5   \\% 61  
38 & 03:18:16.6 & -66:28:45.1 & 0.24 &  	&     &   19.7 &   19.8 &    0.1       & 			 &  			 & 			 &      	     &  	   & NON-WR	\\% 66  
39 & 03:18:17.2 & -66:28:36.5 & 0.28 &  	&     &   20.9 &   20.8 &  --0.1       & 1.7$\times 10^{-16}$	 &   9.6$\times 10^{35}$ & 			 &      	     & WN7--9	       \\% 65  
\hline								
\end{tabular}
\end{table}%
\end{landscape}

\begin{sidewaystable*}
{Table A.2 continued} \\
\begin{tabular}{rrrcccrrrrrrrrl}
\hline \multicolumn{1}{c}{\#}&\multicolumn{1}{c}{$\alpha$} &
\multicolumn{1}{c}{$\delta$}&$\rho/\rho_{0}$& \multicolumn{2}{c}{H\,{\sc ii} Assoc.}& 
m$_{B}$&m$_{4684}$&\multicolumn{1}{c}{$\Delta m$} & F($\lambda$4650)+ &   L($\lambda$4650)+ &  F($\lambda$5808) &L($\lambda$5808)& Sp & Remarks\\
&\multicolumn{2}{c}{(J2000)}&&W\&R&PES&(mag)&(mag)&(mag)&F($\lambda$4686)&L($\lambda$4686)&&&Type\\
\hline 
40 & 03:18:17.6 & -66:30:08.9 & 0.10 &        &     &	24.5 &   22.9 &   -1.6       & 3.1$\times 10^{-16}$	&   1.7$\times 10^{36}$ & 			&		    &  WN2--4	     \\% 53  
41 & 03:18:17.8 & -66:29:03.8 & 0.19 &13      & 9   &	23.2 &   21.0 &   -2.2       & 6.3$\times 10^{-16}$	&   3.5$\times 10^{36}$ & 5.9$\times 10^{-16}$  &2.6$\times 10^{36}$&  WC4--5	     \\% 59  
42 & 03:18:18.2 & -66:29:05.8 & 0.19 &13      & 9   &	\multicolumn{3}{c}{Extended} & 8.7$\times 10^{-16}$	&   4.9$\times 10^{36}$ & 6.1$\times 10^{-16}$  &2.7$\times 10^{36}$&  WC4--5    \\% 58  
43 & 03:18:19.8 & -66:28:43.4 & 0.29 &12      & 11  &	\multicolumn{3}{c}{Extended} &  			&  			& 			&       	    &		  & NON-WR?      \\% 56  
44 & 03:18:20.5 & -66:30:28.6 & 0.22 &        &     &	24.0 &   22.5 &   -1.5       & 1.3$\times 10^{-16}$	&   7.2$\times 10^{35}$ & 			&       	    &  WN2--4	     \\% 33  
45 & 03:18:20.7 & -66:28:29.8 & 0.35 &        &     &	25.0 &   24.0 &   -1.0       & 2.3$\times 10^{-16}$	&   1.3$\times 10^{36}$ & 			&       	    &  WN2--4	     \\% 55  
46 & 03:18:21.0 & -66:28:37.4 & 0.33 &        &     &	21.7 &   21.5 &   -0.2       & 1.3$\times 10^{-16}$	&   7.6$\times 10^{35}$ & 			&       	    &  WN2--4	     \\% 54  
47 & 03:18:21.7 & -66:29:44.7 & 0.21 &        &     &	21.8 &   21.5 &   -0.3       & 7.8$\times 10^{-17}$	&   4.4$\times 10^{35}$ & 			&       	    &  WN7--9	     \\% 36  
48 & 03:18:21.9 & -66:30:06.3 & 0.22 &        &     &	21.1 &   20.0 &   -1.1       & 1.5$\times 10^{-15}$	&   8.6$\times 10^{36}$ & 1.4$\times 10^{-15}$  &6.0$\times 10^{36}$&  WC4--5	 &  2$\times$WC?  \\% 31  
49 & 03:18:21.9 & -66:29:47.9 & 0.22 &        &     &	20.2 &   19.8 &   -0.3       & 6.7$\times 10^{-16}$	&   3.8$\times 10^{36}$ & 4.6$\times 10^{-16}$  &2.0$\times 10^{36}$&  WC4--5	& Multiple  WRs?   \\% 34  
\smallskip50 & 03:18:22.3 & -66:28:42.3 & 0.34 &        & (1) &	22.2 &   21.0 &   -1.2       & 1.1$\times 10^{-15}$	&   6.2$\times 10^{36}$ & 9.8$\times 10^{-16}$  &4.2$\times 10^{36}$&  WC4--5	   \\% 50  
51 & 03:18:22.3 & -66:28:40.5 & 0.35 &        & (1) &  $>$23 &  $>$22 &  $<$-1       & 1.9$\times 10^{-16}$	&   1.1$\times 10^{36}$ & 			&       	    &  WN2--4		   \\% 51  
52 & 03:18:22.5 & -66:30:03.8 & 0.24 &        &     &	22.8 &   21.9 &   -0.9       & 2.3$\times 10^{-16}$	&   1.3$\times 10^{36}$ & 			&       	    &  WN2--4		   \\% 29  
53 & 03:18:22.5 & -66:28:45.6 & 0.34 &        & (1) &$>$23.5 &$>$22.5 &  $<$-1       & 2.4$\times 10^{-16}$	&   1.4$\times 10^{36}$ & 			&       	    &  WN2--4	     \\% 48
54 & 03:18:22.6 & -66:29:41.4 & 0.24 &        &     &	--- &   --- &   ---       &  			&  			& 			&       	    &		  & WN?     \\% 32  
55 & 03:18:22.7 & -66:28:20.2 & 0.41 &        &     &	\multicolumn{3}{c}{Extended} & 1.8$\times 10^{-16}$	&   1.0$\times 10^{36}$ & 			&       	    &  WN2--4	     \\% 52
56 & 03:18:22.8 & -66:28:42.7 & 0.35 &        & (1) &$>$23.5 &    22: & $>$1.5       & 1.7$\times 10^{-16}$	&   9.3$\times 10^{35}$ & 1.3$\times 10^{-16}$  &5.6$\times 10^{35}$&  WC4--5	     \\% 47
57 & 03:18:22.8 & -66:28:36.1 & 0.37 &        & (1) &	21.9 &   21.2 &   -0.7       & 5.6$\times 10^{-16}$	&   3.1$\times 10^{36}$ & 5.5$\times 10^{-16}$  &2.4$\times 10^{36}$&  WC4--5	     \\% 49
58 & 03:18:23.1 & -66:28:40.1 & 0.37 &10      & 1   &	\multicolumn{3}{c}{Extended} & 8.3$\times 10^{-16}$	&   4.7$\times 10^{36}$ & 			&       	    &  WN5--6?	   & 3$\times$WN?  \\% 44
59 & 03:18:23.5 & -66:28:43.4 & 0.37 &10      & 1   &	20.4 &   20.9 &    0.5       &  			&  			& 7.8$\times 10^{-16}$  &3.4$\times 10^{36}$&  WC4--5+WN5--6?    \\% 41   
\smallskip60 & 03:18:23.5 & -66:28:41.4 & 0.37 &10      & 1   &	\multicolumn{3}{c}{Extended} & 4.7$\times 10^{-16}$	&   2.7$\times 10^{36}$ & 8.3$\times 10^{-16}$  &3.7$\times 10^{36}$&  WC4--5    \\% 42
61 & 03:18:23.6 & -66:32:57.9 & 0.74 &28      &     &	21.2 &   20.0 &   -1.2       & \multicolumn{2}{c}{No blue data}			& 2.1$\times 10^{-15}$  &9.4$\times 10^{36}$&  WC4--5	& 3$\times$WC?     \\% 23  
62 & 03:18:23.7 & -66:32:57.7 & 0.74 &28      &     &	21.0 &   20.5 &   -0.5       & \multicolumn{2}{c}{No blue data}			& 1.1$\times 10^{-15}$  &4.6$\times 10^{36}$&  WC4--5	     \\% 22  
63 & 03:18:23.8 & -66:28:21.3 & 0.43 &        &     &	22.8 &   22.3 &   -0.5       &  			&  			& 			&       	    &		  & WN?      \\% 46  
64 & 03:18:23.9 & -66:28:47.2 & 0.37 &10      & 1   &	19.7 &   18.8 &   -0.9       & 4.2$\times 10^{-15}$	&   2.4$\times 10^{37}$ & 2.4$\times 10^{-15}$  &1.1$\times 10^{37}$&  WC4--5	 & 3$\times$WC?     \\% 37  
65 & 03:18:23.9 & -66:28:42.7 & 0.38 &10      & 1   &	\multicolumn{3}{c}{Extended} &  			&  			& 			&       	    &		  & Multiple WN?    \\% 38  
66 & 03:18:23.9 & -66:28:37.9 & 0.39 &10      & 1   &	\multicolumn{3}{c}{Extended} & 1.5$\times 10^{-15}$	&   8.3$\times 10^{36}$ & 1.7$\times 10^{-15}$  &7.7$\times 10^{36}$& WC4--5    &2$\times$WC? \\% 40  
67 & 03:18:24.1 & -66:28:37.3 & 0.40 &10      & 1   &	\multicolumn{3}{c}{Extended} &  			&  			& 1.1$\times 10^{-15}$  &4.9$\times 10^{36}$& WC4--5+WN5--6?   \\% 39  
68 & 03:18:25.7 & -66:28:50.6 & 0.41 &        &     &	24.2 &   23.3 &   -0.9       &  			&  			& 			&       	    &		  & WN?    \\% 30  
69 & 03:18:26.7 & -66:27:56.7 & 0.56 &        &     &	21.4 &   20.6 &   -0.8       &  			&  			& 			&       	    &		  & Late   \\% 35  
\smallskip70 & 03:18:27.1 & -66:28:24.5 & 0.51 &        &     &	23.3 &   22.7 &   -0.6       & 8.4$\times 10^{-17}$	&   4.7$\times 10^{35}$ & 			&       	    &  WN2--4		   \\% 28 
71 & 03:18:27.9 & -66:28:42.8 & 0.49 &9       &     &  $>$23 &   21.7 &$<$-1.3       & 1.9$\times 10^{-16}$	&   1.1$\times 10^{36}$ & 			&       	    &  WN2--4		   \\% 26 
72 & 03:18:27.9 & -66:28:41.9 & 0.49 &9       &     &	20.3 &   20.5 &    0.2       & 7.1$\times 10^{-17}$	&   4.0$\times 10^{35}$ & 			&       	    &  WN7--9		   \\% 27 
73 & 03:18:28.6 & -66:28:48.6 & 0.50 &9       &     &	20.5 &   20.6 &    0.0       & 1.5$\times 10^{-16}$	&   8.6$\times 10^{35}$ & 			&       	    &  WN7--9		   \\% 25 
74 & 03:18:29.2 & -66:28:42.8 & 0.53 &(9)     &     &	21.9 &   22.0 &    0.0       &  			&  			& 			&       	    &  SNR?	  & Neb He\,{\sc ii}	   \\% 24  
75 & 03:18:33.1 & -66:34:04.3 & 1.10 &        &     &$>$24.5 &   23.3 &$<$-1.2       & \multicolumn{2}{c}{No blue data}			& 			&       	    &		  & WN?    \\% 1  
76 & 03:18:34.4 & -66:29:18.2 & 0.65 &        &     &	22.6 &   22.3 &   -0.3       & 2.3$\times 10^{-16}$	&   1.3$\times 10^{36}$ & 			&       	    &  WN2--4		   \\% 20 
77 & 03:18:35.3 & -66:29:21.7 & 0.68 &        &     &	20.6 &   20.4 &   -0.2       & 2.0$\times 10^{-16}$	&   1.1$\times 10^{36}$ & 			&       	    &  WN7--9		   \\% 19 
78 & 03:18:35.8 & -66:29:36.6 & 0.69 &        &     &	22.0 &   22.2 &    0.2       & 2.3$\times 10^{-16}$	&   1.3$\times 10^{36}$ & 			&       	    &  WN2--4		   \\% 18  
79 & 03:18:35.9 & -66:28:38.9 & 0.74 &        &     &	24.6 &   23.4 &   -1.2       & 1.4$\times 10^{-16}$	&   8.1$\times 10^{35}$ & 			&       	    &  WN2--4		   \\% 21  
\hline
\end{tabular}
\end{sidewaystable*}

\begin{sidewaystable*}
{Table A.2 continued} \\
\begin{tabular}{rrrcccrrrrrrrrl}
\hline \multicolumn{1}{c}{\#}&\multicolumn{1}{c}{$\alpha$} &
\multicolumn{1}{c}{$\delta$}&$\rho/\rho_{0}$& \multicolumn{2}{c}{H\,{\sc ii} Assoc.}& 
m$_{B}$&m$_{4684}$&\multicolumn{1}{c}{$\Delta m$} & F($\lambda$4650)+ &   L($\lambda$4650)+ &  F($\lambda$5808) &L($\lambda$5808)& Sp & Remarks\\
&\multicolumn{2}{c}{(J2000)}&&W\&R&PES&(mag)&(mag)&(mag)&F($\lambda$4686)&L($\lambda$4686)&&&Type\\
\hline 

80 & 03:18:36.5 & -66:29:31.0 & 0.72 & 5       & 3   &  25.0 &   24.3 &   -0.7       &  				&		      &  		       &		    &		  & WN?    \\% 16  
81 & 03:18:36.7 & -66:29:32.0 & 0.72 & 5       & 3   & $>$24 &   21.6 &  $<$-2       & 5.4$\times 10^{-18}$		& 3.0$\times 10^{34}$ &   7.7$\times 10^{-16}$ & 3.4$\times 10^{36}$&  WC4--5		   \\% 12  
82 & 03:18:36.9 & -66:29:24.1 & 0.73 &         &     & $>$24 &$>$23.5 &$<$-0.5       & \multicolumn{2}{c}{No blue data}  		      &	        	       &		    &		  & WN?    \\% 13  
83 & 03:18:36.9 & -66:29:15.1 & 0.74 &         &     &  21.8 &   21.5 &   -0.4       & 1.1$\times 10^{-16}$		& 6.3$\times 10^{35}$ &  		       &		    &  WN7--9		   \\% 17  
84 & 03:18:37.1 & -66:29:18.9 & 0.74 &         &     &  20.4 &   20.2 &   -0.2       &  				&		      &  		       &		    &		  & NON-WR?    \\% 14  
85 & 03:18:37.5 & -66:29:34.8 & 0.75 & 5       & 3   &  19.0 &   19.9 &    0.9       & 2.4$\times 10^{-16}$		& 1.4$\times 10^{36}$ &  		       &		    &  WN7--9	  & Multiple  WRs?	   \\% 10  
86 & 03:18:38.4 & -66:28:46.5 & 0.81 & (6)     &     &  22.2 &   21.5 &   -0.6       & \multicolumn{2}{c}{No blue data}	 		      &	        	       &		    &		  & WN?    \\% 15  
87 & 03:18:38.6 & -66:29:09.3 & 0.80 &         &     & $>$24 &   23.9 &  $<$-0.1     &  				&		      &  		       &		    &		  & NON-WN?    \\% 11  
88 & 03:18:41.0 & -66:30:08.9 & 0.87 & 4       &     &  19.8 &   19.7 &   -0.1       & \multicolumn{2}{c}{No blue data} 		      & 1.4$\times 10^{-15}$   & 6.1$\times 10^{36}$&  WC4--5	  & 2$\times$WC? \\% 5/6 
89 & 03:18:41.5 & -66:30:19.7 & 0.89 &         &     &  21.3 &   21.4 &    0.1       &  				&		      &  		       &		    &		  & NON-WR?    \\% 2   
\smallskip90 & 03:18:41.9 & -66:29:31.0 & 0.90 & 3       & 10  &  19.5 &   19.3 &   -0.2       & 1.8$\times 10^{-16}$		& 1.0$\times 10^{36}$ &  		       &		    &  WN7--9	  & Multiple  WRs?	   \\% 9   
91 & 03:18:42.0 & -66:29:41.9 & 0.90 &         &     &  22.1 &   21.9 &   -0.1       & 9.0$\times 10^{-17}$		& 5.1$\times 10^{35}$ &  		       &		    &  WN7--9		   \\% 7   
92 & 03:18:42.4 & -66:29:33.0 & 0.92 & 3       & 10  &  20.9 &   21.0 &    0.1       & 1.2$\times 10^{-16}$		& 6.7$\times 10^{35}$ &  		       &		    &  WN7--9		   \\% 8   
93 & 03:18:42.5 & -66:29:41.4 & 0.92 &         &     &  20.7 &   20.7 &    0.0       &  				&		      &  		       &		    &		  & NON-WR?   \\% 4   
94 & 03:18:44.5 & -66:29:08.5 & 1.00 &         &     &  22.2 &   21.2 &   -1.0       & 1.1$\times 10^{-16}$		& 6.3$\times 10^{35}$ &  		       &		    &  WN7--9		   \\% 3 
\hline
\end{tabular}
\end{sidewaystable*}%

\begin{figure*}
\centerline{\psfig{figure=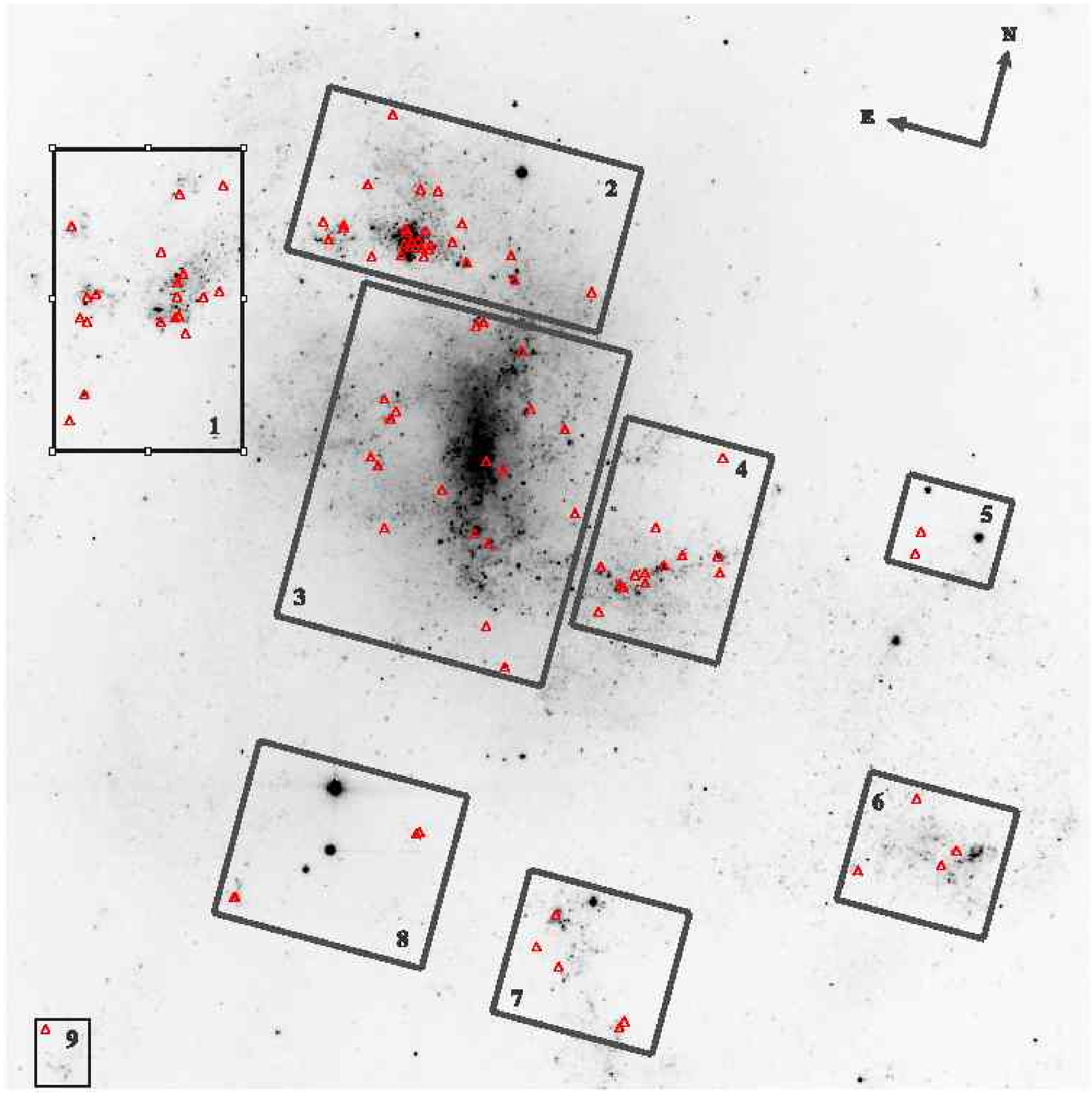,width=13cm,angle=0.}}
\begin{flushleft}
{Fig. B.1: Master finding chart indicating the 94 candidates
identified (triangles) and positions of individual finding
charts~(1--9).  North and east are marked on this $\lambda$4684 FORS1
image of NGC\,1313 (6.8 $\times$ 6.8 arcmin).  }
\end{flushleft}

\end{figure*}

\end{document}